\documentclass[12pt,preprint]{aastex}

\usepackage{natbib}

\slugcomment{Not to appear in Nonlearned J., 45.}

\shorttitle{Star formation from UV and IR points of view}
\shortauthors{Iglesias-P\'{a}ramo et al.}

\begin{document}

\title{Star formation in the nearby universe: the ultraviolet and infrared points of view}


\author{J. Iglesias-P\'{a}ramo\altaffilmark{1,10}, V. Buat\altaffilmark{1}, T. T. Takeuchi\altaffilmark{1}, K. Xu\altaffilmark{2}, 
S. Boissier\altaffilmark{3}, A. Boselli\altaffilmark{1}, D. Burgarella\altaffilmark{1}, B. F. Madore\altaffilmark{3}, A. Gil de Paz\altaffilmark{3}, L. Bianchi\altaffilmark{4}, T. A. Barlow\altaffilmark{2}, 
Y.-I. Byun\altaffilmark{5}, J. Donas\altaffilmark{1}, K. Forster\altaffilmark{2}, P.G. Friedman\altaffilmark{2}, T. M. Heckman\altaffilmark{6}, P. N. Jelinski\altaffilmark{7}, Y.-W. Lee\altaffilmark{5}, 
R. F. Malina\altaffilmark{1}, D. C. Martin\altaffilmark{2}, B. Milliard\altaffilmark{1}, P. F. Morrissey\altaffilmark{2}, S. G. Neff\altaffilmark{8}, R. M. Rich\altaffilmark{9}, D. Schiminovich\altaffilmark{2},
M. Seibert\altaffilmark{2}, O. H. W. Siegmund\altaffilmark{7}, T. Small\altaffilmark{2}, A. S. Szalay\altaffilmark{6}, B. Y. Welsh\altaffilmark{7}
}

\and

\author{T. K. Wyder\altaffilmark{2}}

\altaffiltext{1}{
Laboratoire d'Astrophysique de Marseille, 13376 Marseille, FRANCE}
\altaffiltext{2}{
Space Astrophysics Laboratory, Mail Stop 405-47, California Institute of Technology, 1200 East California Boulevard, Pasadena, CA 91125}
\altaffiltext{3}{
Observatories of the Carnegie Institution of Washington, 813 Santa Barbara Street, Pasadena, CA 91101}
\altaffiltext{4}{
Center for Astrophysical Sciences, Johns Hopkins University, 3400 North Charles Street, Baltimore, MD 21218}
\altaffiltext{5}{
Center for Space Astrophysics, Yonsei University, Seoul 120-749, Korea}
\altaffiltext{6}{
Department of Physics and Astronomy, Johns Hopkins University, Homewood Campus, Baltimore, MD 21218}
\altaffiltext{7}{
Space Sciences Laboratory, University of California at Berkeley, 601 Campbell Hall, Berkeley, CA 94720}
\altaffiltext{8}{
Laboratory for Astronomy and Solar Physics, NASA Goddard Space Flight Center, Greenbelt, MD 20771}
\altaffiltext{9}{
Department of Physics and Astronomy, University of California, Los Angeles, CA 90095}
\altaffiltext{10}{
Instituto de Astrof\'{\i}sica de Andaluc\'{\i}a, 18008 Granada, SPAIN}




\begin{abstract}
This work presents the main ultraviolet (UV) and far-infrared (FIR) properties of two samples of nearby galaxies selected from the 
GALEX ($\lambda = 2315$\AA, hereafter NUV) and IRAS ($\lambda = 60\mu$m) surveys respectively. They are built in order to get detection at both wavelengths for most of the galaxies. Star formation rate (SFR)
estimators based on the UV and FIR emissions are compared.
Systematic differences are found between the SFR estimators for individual galaxies based on the NUV fluxes corrected for dust attenuation and on the total IR luminosity. A combined estimator based on NUV and IR luminosities seems to be the best proxy over the whole range of values of SFR. 
Although both samples present similar average values of the birthrate parameter b, their star-formation-related properties are substantially different: NUV-selected galaxies tend to show larger values of $b$ for lower masses, SFRs and dust attenuations, supporting previous scenarios for the star formation history (SFH). Conversely, about 20\%
of the FIR-selected galaxies show high values of $b$, SFR and NUV attenuation. These galaxies, most of them being LIRGs and ULIRGs, break down the downsizing picture for the SFH, however their relative contribution per unit volume is  small in the local Universe.
Finally, the cosmic SFR density of the local Universe is estimated in a consistent way from the NUV and IR luminosities.
\end{abstract}



\keywords{surveys: GALEX --- ultraviolet: galaxies --- infrared: galaxies}



\section{Introduction}

What is the best way to measure the SFR of galaxies on large scales
and at different redshifts?  The possibility of estimating the SFR of
a galaxy directly from the luminosity at a single wavelength would be
a major advantage for anyone wanting to compute the SFR per unit
volume at a given redshift. This quantity could be derived directly
from the luminosity function (LF) at this wavelength and at this
redshift. Under these conditions, large area surveys at single
wavelengths might suffice.

The recent SFR of a galaxy is often measured from the light emitted by the young stars: given their short lifetimes their luminosity is directly proportional to the rate at which they are currently forming.
The UV and FIR luminosities of star forming galaxies are both closely related to the recent star formation: most of the UV photons are originally emitted by stars younger than $\sim 10^{8}$~yr, but many of these photons are re-processed by the dust present in galaxies and re-emitted at FIR wavelengths.
Strictly speaking, neither of these fluxes can be used alone to estimate the SFR independently of the other one (e.g. Buat \& Xu 1996; Hirashita et al. 2003; Iglesias-P\'{a}ramo et al. 2004). Because of the previous lack of data at both wavelengths, attempts have been made using only the rest frame UV 
(Lilly et al. 1996; Madau et al. 1996; Steidel et al. 1999; and more recently Schiminovich et al. 2005 with GALEX data), or just FIR data (Rowan-Robinson et al. 1997; Chary \& Elbaz 2001), but only a few authors have compared both (Flores et al. 1999, Cardiel et al. 2003).
The SFR estimator based on the UV luminosity suffers from attenuation by dust and it has to be corrected  in order to properly trace the SFR: for UV-selected samples of galaxies the attenuation can reach more than 1~mag (e.g. Iglesias-P\'{a}ramo et al. 2004; Buat et al. 2005). On the other hand the FIR emission is not free of problems because the dust can also be heated by old stars, and can be a non-negligible correction for many star forming galaxies (Lonsdale \& Helou 1987; Sauvage \& Thuan 1992).
Neither of these two indicators taken alone is an accurate estimator of the SFR except perhaps for starburst galaxies where (a) the dust attenuation was found to follow a tight relation with the slope of the spectrum at UV wavelengths (Meurer et al. 1999), thus allowing one to estimate the dust attenuation with only information on UV fluxes, and (b) the contribution to the dust emission coming from old stars can be neglected (Sauvage \& Thuan 1992). In the most general case, the best estimator of the SFR should contain combined information of the luminosities at both wavelength ranges (Hirashita et al. 2003).
The UV and FIR fluxes are thus complementary for tracing star formation and it is well known that the FIR/UV ratio is a proper indicator of the dust attenuation (Buat et al. 1999; Witt \& Gordon 2000; Panuzzo et al. 2003). 
Although other indicators of the recent SFR of galaxies have been extensively used in the literature, a detailed discussion of their quality as SFR tracers will not be discussed in this paper.

The GALEX mission (Martin et al. 2005a) is imaging the high-Galactic
latitude sky at two UV wavelengths ($\lambda=1530$\AA, FUV;
$\lambda=2315$\AA, NUV) and is providing the astronomical community
with unprecedented data (both in quantity and quality). The UV data
combined with existing FIR datasets (from the IRAS, ISO or
Spitzer missions) now allow us to carry out detailed studies of the UV and
FIR properties of galaxies, with special emphasis on the derivation of
the dust attenuation and star formation activity in star forming
galaxies.

With this purpose in mind we selected two samples of galaxies: one from the GALEX-All Imaging Sky (AIS) and the other from the IRAS PSCz (redshifts, infrared and optical photometry, and additional information for 18,351 IRAS sources, mostly selected from the Point Source Catalog) and FSC (Faint Source Catalog)
for which UV and FIR fluxes are available. With these datasets in hand we undertaken a study of the properties related to their emission at these wavelengths. 
Both samples were extracted from the same region of the sky ($\sim 600$~degrees$^{2}$, constrained by the status of the GALEX survey when this work was initiated). From the GALEX catalog we built a complete sample of galaxies down to $AB_{NUV}=16$~mag\footnote{AB magnitudes are defined as: $AB_{\nu} = -2.5 \log f_{\nu} -48.6$, where $f_{\nu}$ is the monochromatic flux density expressed in erg~s$^{-1}$~cm$^{-2}$~Hz$^{-1}$.} and cross-correlated it with the IRAS database (FSC), allowing non-detections at 60$\mu$m (fluxes lower than 0.2 Jy).
The FIR-selected sample was built from the IRAS catalog (PSCz) with a limiting flux at 60$\mu$m of 0.6~Jy as the only constraint. The resulting list of PSCz sources was cross-correlated with the GALEX database, 
allowing again non-detections in the NUV. 
Both the NUV and the 60$\mu$m limits used to build the samples were chosen to allow for a very small number of non-detections  and to  sample the galaxy population over a large range of values of the dust attenuation. Besides an  analysis of the SFR, for  these
 NUV and FIR-selected samples of galaxies (chosen with well-defined selection criteria) can also be used to place  important constraints on  models designed to predict the statistical properties of galaxy populations.
The attenuation of star light by interstellar dust, and its
emission in the far infrared are usually computed very crudely in
models of galactic evolution. Dust attenuation is usually deduced in such
approaches from other quantities such as the mass gas and the
metallicity (e.g. Guiderdonni \& Rocca-Volmerange 1987; Devriendt \& Guiderdoni 2000; Balland et al. 2003). 
The properties of large samples of galaxies observed both
in the UV and FIR with clear selection criteria, such as the ones presented
in this paper, provide an important statistical constraint for the
calibration of the treatment of dust in such models.

The first paper in this series (Buat et al. 2005) based on these samples was mainly devoted to the dust attenuation properties. In the present work we discuss various aspects  relating to the NUV and 60$\mu$m emission of our sample galaxies, including systematic differences in the 60$\mu$m and NUV luminosities and we will focus on their star formation related properties.
This paper is organized as follows: 
the samples are presented in section~2. The relation between NUV and 60$\mu$m luminosities is discussed in section~3. Section~4 is devoted to the derivation of the SFR and to  a comparison of various estimators as well as to a discussion on the star formation activity related properties of the samples. 
The derivation of the local cosmic SFR density by different methods is discussed in section~5.
The main conclusions are presented in section~6.
Throughout this paper we adopt the following cosmological parameter set: $(h, \Omega_0, \lambda_0)=(0.7,0.3,0.7)$, where 
$h\equiv H_0/100\; (\mbox{km\,s}^{-1}\mbox{Mpc}^{-1})$.

\section{Observational dataset}

\subsection{NUV-selected sample \label{uvsel_sam}}

In order to build the NUV-selected sample, we used 833 fields of the GALEX All sky Imaging Survey 
(AIS) each having an exposure time equal to or larger than 50s. We used only the central 0.5~deg radius 
circles in each field to ensure a uniform image quality: the resulting sky coverage is 
615 deg $^2$. Within this area we selected all the galaxies of the GALEX AIS survey with
$AB_{NUV} \leq 16$~mag. 
This bright limit was chosen in order to ensure IRAS detections of all the galaxies with 
attenuation larger than $\sim 0.3$~mag (for a limit of 0.2 Jy at 60 $\mu$m from the IRAS FSC using
the calibration of Buat et al. 2005) and highly significant upper limits for the less attenuated galaxies.
 
On the one hand, the moderate angular resolution of GALEX (FWHM $\sim 6$~arcsecs) does not allow 
for a secure discrimination of stars from small galaxies.
On the other hand the GALEX pipeline can induce some shredding of larger galaxies. This results in multiple detections which (cumulatively) correspond to a single galaxy because of the misidentification of star forming regions as if they were individual galaxies. The main consequence of this is the underestimation of the fluxes of large galaxies.
Corollary catalogs were thus required in order to perform a reliable selection of galaxies.
Our starting point was the catalog of NUV detections produced by the GALEX pipeline\footnote{version 0.2.0, September 2003, with the correction to the NUV and FUV magnitudes reported in Buat et al. 2005}, which made use of the Sextractor code (Bertin \& Arnouts 1996). We made the assumption that all the galaxies brighter than $AB_{NUV}=$16~mag, even if they are shredded,
should contain at least one Sextractor detection brighter than $AB_{NUV}=18$~mag. 
Then we associated an object from databases of well known stars and galaxies (SIMBAD, 2MASS, LEDA) with each of the Sextractor detections brighter than $AB_{NUV}=18$~mag.
The problem of shredding was mostly resolved by using LEDA. As this database contains the optical diameters of the galaxies, NUV detections corresponding to shredded galaxies can be associated with
their parent galaxies provided they are located within the aperture defined by the optical diameters and the position angle listed in LEDA.
For the detections of galaxies not shredded we used SIMBAD and 2MASS
in order to classify them as stars or galaxies. Finally, objects not associated with any known source were classified as ``dubious''.
In order to test the quality of our selection method, we cross-correlated our final catalog with the SDSS DR1, which spatially overlap one fifth of our sample. The spectral and photometric information of the SDSS together with its higher angular
resolution made possible an optimal classification of all the objects detected in the field:
 all the objects present in both GALEX and SDSS  catalogs were found to be  properly
classified.
Dubious objects were found to be noise detections or ghosts generated mainly near the edges of the GALEX frames.
Thus we ended up with a catalog of {\em bona fide} galaxies or fragments of galaxies (i.e. belonging to the large galaxies shredded by Sextractor) brighter than $AB_{NUV}=18$~mag. The next step was to obtain aperture photometry of these objects in order to select out only the galaxies brighter than $AB_{NUV}=16$~mag.

Photometry of our sample galaxies was performed in the GALEX NUV and FUV bands. Since the selection criterion for our sample galaxies was imposed in the NUV band, we also took the NUV images as our reference for the
total photometry. 
We performed aperture photometry using a set of elliptical apertures, the total photometry corresponding to the aperture where convergence of the growth curve is achieved.
Once we determined the total NUV flux, the photometry in the FUV band was obtained by performing background subtracted aperture photometry within the same elliptical aperture where convergence of
the NUV growth curve was achieved. This way, the NUV$-$FUV colors are consistent.
Some galaxies were contaminated by the flux of nearby bright stars or galaxies. The contaminating sources were then masked in the NUV and FUV frames in order to obtain proper NUV and FUV fluxes for our galaxies.
Table~\ref{error} shows the typical uncertainties of the NUV and FUV magnitudes.
NUV and FUV magnitudes were corrected for Galactic extinction using the Schlegel et 
al. (1998) dust map and the Cardelli et al. (1989) extinction curve.
In the end, a total of 95 non-stellar objects brighter than $AB_{NUV}=16$~mag were found. One of them, classified as a QSO, was excluded of the sample.

FIR fluxes at 60$\mu$m were obtained from the IRAS Faint Source Catalog (FSC, Moshir et al. 2000) for 68 of our 95
galaxies. 
We discarded all these sources for which the cirrus parameter (as listed by the FSC) is larger than 2 because
it can result in uncertain fluxes.
The Scan Processing and Integration Facility (SCANPI) was used to obtain the FIR 
fluxes for the remaining 27 objects. Three of these galaxies (UGC~11866, UGCA~438 and UGC~12613) were not detected at 60$\mu$m. 
We adopted a conservative upper limit of 0.2Jy at 60$\mu$m (as given in the FSC) for these galaxies.
Four galaxies of the sample were not covered by the IRAS survey.

\subsection{FIR-selected sample \label{firsel_sam}}

The FIR-selected sample was extracted from the IRAS PSCz (Saunders et al. 2000)
over the 509~$\deg^{2}$ in common with the GALEX AIS fields having exposure 
times larger than 90~sec. 
In order to keep only good quality FIR data we discarded those galaxies 
for which the probability of a correct optical identification 
of the FIR-selected galaxies was lower than 50\%, as listed in the PSCz.
As for the NUV-selected sample, galaxies for which the cirrus parameter (as listed in the PSCz)
was larger than 2 were discarded.
A total of 163 galaxies were selected; all but two of them (Q00443+1038 and Q23367-0448) 
have published radial velocities.
As galaxies were selected from the PSCz, the imposed limiting flux at 60$\mu$m was 0.6~Jy.
This limit, combined with the one estimated for the GALEX AIS at NUV ($AB_{NUV} \sim 20.5$~mag, Morrissey et al. 2005), results in detections at NUV for all the galaxies with dust attenuation as large as $\sim 4.4$~mag (Buat et al. 2005). Indeed, only two galaxies (Q00443+1038 and Q00544+0347) 
were not detected in the NUV frames and a total of 23 were not detected in the FUV frames.

The NUV and FUV photometry of the FIR-selected galaxies was performed using 
the same technique as for the NUV-selected galaxies.

\subsection{Completeness and bias of the samples}

Before drawing conclusions about the properties of the samples we have to check on just how representative they are. Because of the reduced statistics of the samples, if the sampled volumes are not large enough it could be that some luminosity ranges are oversampled or undersampled with respect to reference samples defined over larger volumes of the local Universe.

\begin{figure}
\epsscale{1.00}
\plotone{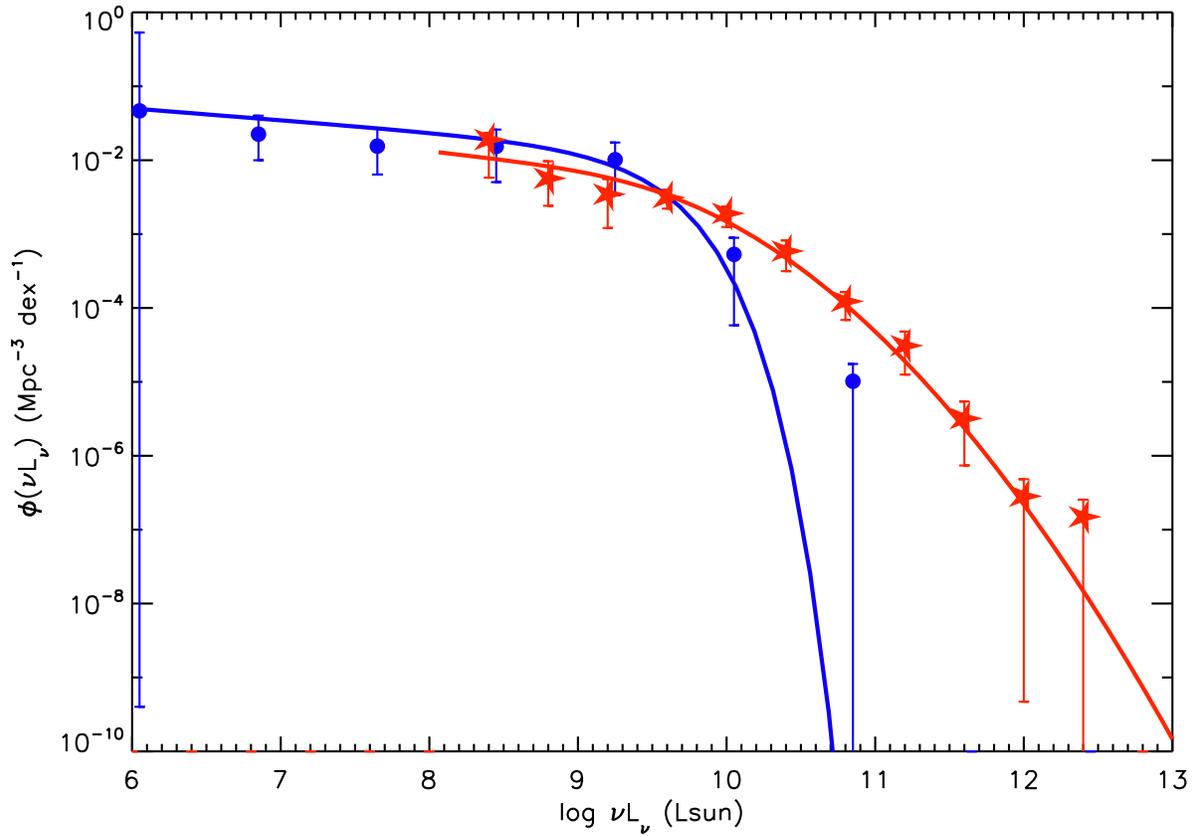}
\caption{
NUV (blue filled dots) and 60$\mu$m (red stars) LFs for the NUV and FIR-selected samples. The blue and red lines correspond to the NUV and 60$\mu$m LFs from Wyder et al. (2005) and Takeuchi et al. (2003) respectively. Error bars correspond to 1-$\sigma$ uncertainty.
}
\label{nlf}
\end{figure}

We check the representative nature of our samples by building  LFs and comparing them to the standard ones at $z=0$, constructed from larger samples of galaxies (NUV LF of Wyder et al. (2005) and 60$\mu$m LF of Takeuchi et al. (2003)). 
We calculate both NUV and 60$\mu$m LFs of our sample by the Lynden-Bell method 
(Lynden-Bell 1971), implemented to obtain the normalization using the 
formulation of Takeuchi et al. (2000).
The calculation of the uncertainty is based on a bootstrap resampling method (Takeuchi et al.~(2000)).
We note that the Lynden-Bell method is robust against  density 
inhomogeneities, and hence we can trust the LF so determined (see Takeuchi et al.~2000 for details).
The results are shown in Figure~\ref{nlf}. 
The error bars correspond to 1$\sigma$ uncertainties.
The agreement between the LFs of our samples and the corresponding LFs from
larger samples is very good, so we are confident that in spite of their 
small size our samples are representative of flux-limited NUV and FIR 
samples in the local Universe.

\begin{figure}
\epsscale{0.80}
\plotone{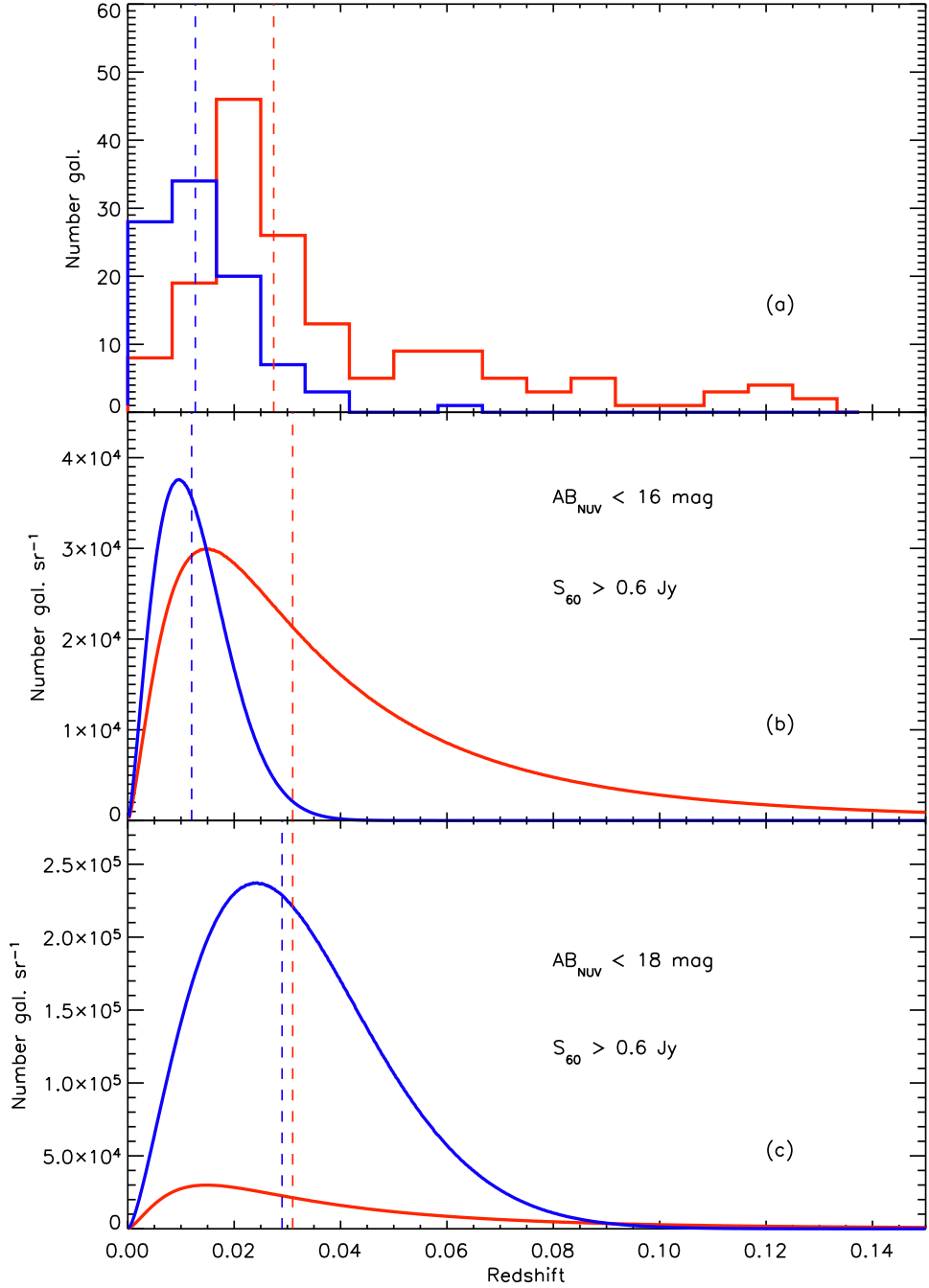}
\caption{
(a) Redshift distributions for the NUV (blue) and FIR (red) selected galaxies. (b) Theoretical redshift distributions for samples of galaxies with the same limiting fluxes as our samples. (c) Same as (b) but for a NUV-selected sample with $AB_{NUV} \leq 18$~mag. Vertical dashed lines in the three panels correspond to the median values of each distribution. 
\label{histo_velo_all}
}
\end{figure}

We also compare the volumes from which the samples were extracted. 
Figure~\ref{histo_velo_all}a shows the redshift distributions 
for both samples. 
The median values are 0.013 and 0.027 for the NUV and FIR-selected 
samples respectively. 
At a first glance, this means that the FIR selection samples a volume 8 
times larger than the corresponding NUV selection. 
However, we must recall that the redshift distribution of a flux limited 
sample is strongly dependent on the shape of the LF, 
and as shown in Figure~\ref{nlf}, the NUV and 60$\mu$m LFs are very different. 
In Figure~\ref{histo_velo_all}b we show the theoretical redshift distributions 
for NUV and FIR-selected samples with the same limiting fluxes as our two 
samples\footnote{Details on the calculation are given in 
Appendix~\ref{depth}.}.
As can be seen, the theoretical median values of the redshift for both samples are consistent with the ones obtained  from the observational data.
A limiting magnitude of $AB_{NUV} = 18$~mag is required to obtain similar median values of the redshift distributions of both samples, as we show in 
Figure~\ref{histo_velo_all}c. 
And in this case most of the galaxies detected in NUV will not have any 
counterpart in FIR. 
This behavior of the redshift distribution can be understood intuitively. 
Indeed, the flux density selection procedure omits intrinsically 
low-luminosity objects from the sample, whereas bright objects are hardly 
affected by the flux selection. 
To the degree that the LF is well reconstructed from the flux/magnitude-limited
samples, these samples can be said to be representative, with respect to 
the luminosity and/or flux density, and it is indeed the case for 
the present work.

\section{Relation between $L_{60}$ and $L_{NUV}$ \label{tsu_vero}}

The physical link between the FIR and UV luminosities of galaxies is rather 
complex. On the one hand, both are related to the light of young stars, so one 
expects a correspondence between the two. On the other hand, the FIR emission is due to 
the absorption of the UV light thereby leading to an anti-correlation. Since our samples were built to avoid upper limits -- i.e. most of the galaxies selected in NUV (or at 60 $\mu$m) are also detected at 60 $\mu$m (or in NUV) -- we are able to discuss statistically the intrinsic relation between both the two wavelengths and  outline the specifics of  NUV versus  FIR selection effects.

\begin{figure}
\epsscale{1.00}
\plotone{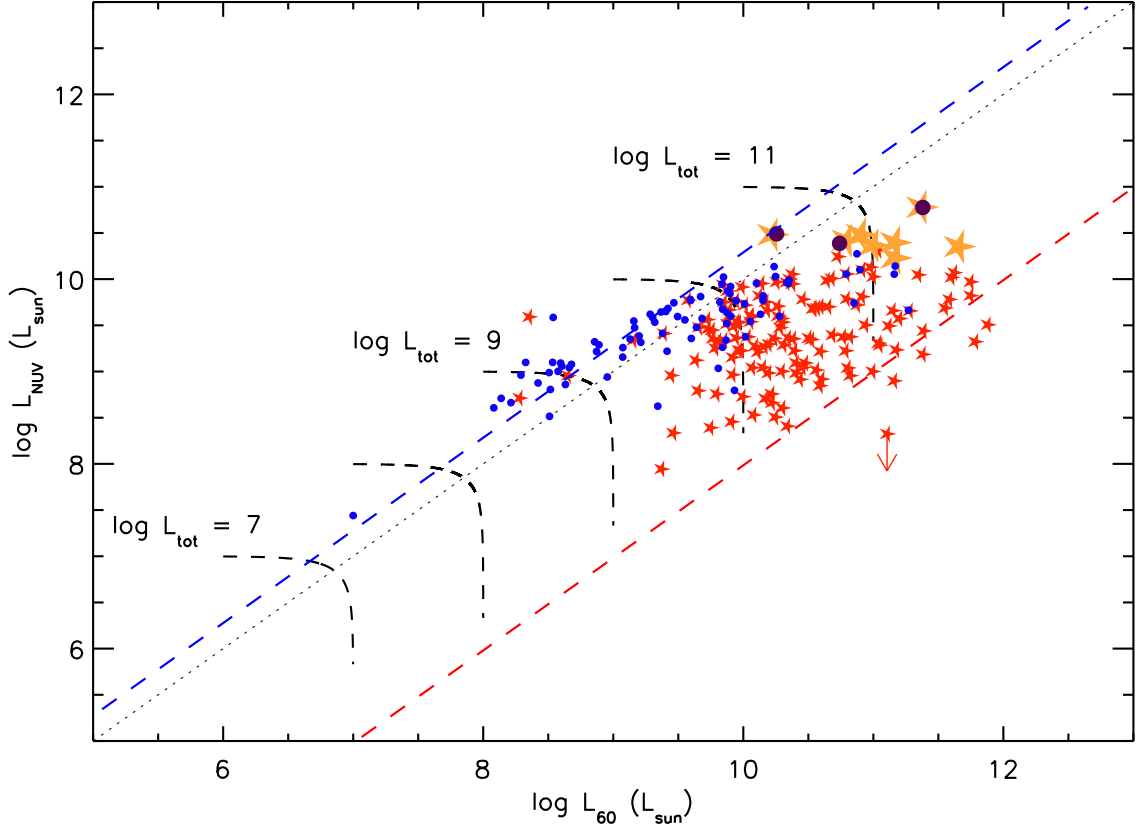}
\caption{$L_{60}$ vs. $L_{NUV}$ (non corrected for attenuation) for the NUV (blue circles) and FIR (red stars) selected samples. Arrows indicate upper limits. Dashed curves represent the loci of points with $L_{tot} = 10^7, 10^8, 10^9, 10^{10}$ and $10^{11}L_{\odot}$ respectively. The dotted line corresponds to $L_{NUV} = L_{60}$. The red (blue) dashed line corresponds to the maximum (minimum) attenuation below (above) which the FIR (NUV) selected sample is complete. The larger stars and circles correspond to the UVLGs of the FIR and NUV selected samples respectively.
\label{l60_lnuv}}
\end{figure}

In Figure~\ref{l60_lnuv} we plot $L_{NUV}$ versus $L_{60}$ for both samples
\footnote{Throughout the paper the NUV and 60 $\mu$m luminosities $L_{NUV}$ and $L_{60}$ will be calculated  as  $\nu L_{\nu}$ expressed in 
solar units. The adopted value for the bolometric solar luminosity is $L_{\odot} = 3.83 \times 10^{33}$~erg~s$^{-1}$.}.
The two samples exhibit very different behaviors: 
the NUV-selected galaxies show a good correlation between both luminosities,
with a correlation coefficient (in log units) of $r
\simeq 0.8$. On the contrary
the dispersion is very large for FIR-selected galaxies and the
correlation coefficient is low: $r \simeq 0.3$. 

NUV-selected galaxies appear intrinsically less luminous
at 60$\mu$m than FIR-selected ones. 
This is also true for the sum of both luminosities, $L_{tot} = L_{NUV} +
L_{60}$, which is supposed to be a crude estimate of the bolometric luminosity of galaxies 
related to recent star formation (e.g. Martin et al. 2005b). Although the luminosity 
distribution within each sample is the combined effect of the LFs 
and  selection criteria, 
this result is confirmed by other studies and from the comparison of the 60$\mu$m and NUV LFs themselves (e.g. Martin et al. 2005b; Buat et al. 2005).
Both distributions flatten at higher luminosities, reflecting a general increase of 
the dust attenuation already pointed out in the literature by several authors (Wang \& Heckman 1996; Buat et al. 1999; Sullivan et al. 2001; Vijh et al. 2003).

One could argue that the difference in luminosity between the two samples is a consequence of bias in the  sampling. 
We show in Figure~\ref{l60_lnuv} the lines corresponding to our lower (upper) limit of the NUV attenuation above (below) which the NUV (FIR) selected sample is complete. Thus, the fact that only very few low-luminosity and low-attenuation FIR-selected galaxies are detected must be taken as real. Low-luminosity, high-attenuation galaxies should have been detected by our FIR survey if they were present. For the same reason, very luminous galaxies should have been detected in the NUV survey if they existed. 

The good correlation found between $L_{NUV}$ and $L_{60}$ for the NUV-selected galaxies has to be related to their rather low dust attenuation: in these galaxies both $L_{NUV}$ and $L_{60}$ represent a significant part of the total luminosity of the galaxies. This result holds for intrinsically faint galaxies ($L_{tot} \leq 2 \times 10^{9} L_{\odot}$). The very loose correlation found for FIR-selected galaxies may  also be explained by the effect of the dust attenuation. With a mean dust attenuation larger than 2~mag, the NUV luminosity becomes a  poor tracer of their total 
luminosity whereas the 60$\mu$m luminosity is not very different from the bolometric emission of the young stars.
Some fluctuations in the percentage of  NUV photons escaping the galaxies can induce large variations in the NUV 
observed luminosity on an absolute scale without any strong physical 
difference on the scale of the total luminosity of the galaxies.

We make a final comment on the so-called ``UV luminous galaxies'' (UVLGs, defined as those with $L_{FUV} \geq 2 \times 10^{10}$~L$_{\odot}$ in Heckman et al. 2005). We found 3 ULVGs in our NUV-selected sample and a total of 8 (including the previous 3) in the FIR-selected sample. All but one of these galaxies are more luminous at 60$\mu$m than in the  NUV (in fact most of them are LIRGs), and their attenuation is typically larger than 1~mag. This means that these galaxies are not only UV luminous but also very luminous from a bolometric point of view.

\section{Selection effects on observational quantities and physical properties of galaxies}

The main aim of this section is to show the effect of the selection criteria of samples of galaxies on  observational and physical quantities. We will now show that the selection criteria of a sample of galaxies play an important role in defining the nature of the galaxies selected and thus, in their averaged properties. Accordingly we warn against the unqualified comparison of results obtained from samples of galaxies selected on the basis of different criteria. 

In order to reduce 
the uncertainties associated with the FIR and NUV fluxes we  
impose further constraints on our  galaxy sample:
\begin{itemize}
\item Ellipticals, S0s as well as AGNs (Seyferts and QSOs) were excluded since the origin 
of their 60$\mu$m and NUV fluxes is clearly not associated to recent star formation. 
The necessdary classification information is available for most of the NUV-selected galaxies; but this turned 
out not to be the case for the FIR-selected galaxies, so contamination of the sample by ellipticals 
and/or AGN among these galaxies cannot be totally excluded.
Galaxies with extraneous radio sources (from NVSS and/or FIRST) within the IRAS beam 
were also excluded since part of the FIR flux of these galaxies could be due to 
contaminating background objects. 
\item Multiple galaxies, not resolved by the IRAS 
beam but clearly resolved into various components in the GALEX frames were excluded, since a one-to-one 
60$\mu$m-NUV association is not possible for them.

\end{itemize}

After applying these criteria we ended up with 59 and 116 galaxies from the original NUV and FIR-selected samples, respectively. 
Hereafter we will use these restricted subsamples for our subsequent analysis of the star formation related properties, although we will keep the terminology FIR and NUV-selected samples to refer to the restricted subsamples.
Given that all the galaxies were extracted from the same region of the sky, 
some of them belong to both subsamples. Their basic properties are listed in Table~2: (1) Identifier of the galaxy; (2) ``Y'' (``N'') indicates whether the galaxies is included (or not) in the NUV-selected sample; (3) ``Y'' (``N'') indicates whether the galaxies is included (or not) in the FIR-selected sample; (4) R.A.(J2000 equinox) of the source ; (5) Declination (J2000 equinox); (6) Radial velocity in km~s$^{-1}$, obtained from NED or LEDA; (7) Distance to the source in Mpc, corrected for the Local Group Infall to Virgo and $H_{0} = 70$~km~s$^{-1}$~Mpc$^{-1}$; (8) Morphological type, from NED or LEDA; (9) IRAS identifier: ``F'' for FSC; ``Q'', ``O'' or ``R'' for PSCz; ``SCANPI'' for absence in both catalogs.

Table~\ref{photo} gives some useful photometric data for the galaxies in the restricted subsamples: (1) Optical identifier; (2) NUV magnitude corrected for Galactic extinction; (3) FUV magnitude corrected for Galactic extinction; (4) Flux density at 60$\mu$m in Jy; (5) Flux density at 100$\mu$m in Jy; (6) $H$ magnitude from 2MASS Extended Source Catalog. For galaxies with no detection by 2MASS we adopt the limiting value of $H = 13.9$~mag (3~mJy) as given by Jarrett et al. (2000); (7) NUV attenuation in mag, derived as in Buat et al. (2005); (8) FUV attenuation derived as indicated in Buat et al. (2005). 
For some galaxies Eqs.~\ref{attenuv_eq} and \ref{attefuv_eq} gave negative values of the NUV and FUV attenuations which is, of course, unphysical. In fact, this is an artifact of the polynomial fitting used to derive a functional form for the attenuation in Buat et al. (2005). Throughout this paper they will be considered as zero.
Given that the FIR fluxes were extracted from different catalogs (PSCz for the FIR-selected sample and FSC/SCANPI for the NUV-selected sample), for those galaxies belonging 
to both samples we list the FIR entries corresponding to the PSCz catalog. 
For those galaxies not present in the FSC and not detected by SCANPI at 60$\mu$m, we list 
$f_{60} \leq 0.2$Jy, which is the nominal limiting flux of the FSC.  No estimate of an upper 
limit at 100 $\mu$m is given for these galaxies.
For galaxies with no detection in 2MASS we adopt the limiting value of $H = 13.9$~mag (3~mJy) as published by Jarrett et al. (2000).

In Table~\ref{sfr_tab} we list some star-formation properties which will be used in the forthcoming discussion: (1) Identifier of the galaxy; (2) $SFR_{NUV}$ from Eq.~\ref{sfrnuv_eq}; (3) $SFR_{FUV}$ from Eq.~\ref{sfrfuv_eq}; (4) $SFR_{dust}$ from Eq.~\ref{sfrfir_eq}; (5) $SFR_{tot}(NUV)$ from Eq.~\ref{sfrtot_eq}; (6) $SFR_{tot}(FUV)$ from Eq.~\ref{sfrtot_eq} but modified by using $SFR^{0}_{FUV}$ instead of $SFR^{0}_{NUV}$; (7) $\left< SFR \right>$ averaged over the galaxy's lifetime, estimated as indicated in Appendix~\ref{apendice_b}.

\subsection{SFR derivations}

This section is devoted to a detailed comparison of the recent SFR as seen in the FIR and NUV-selected samples.
Although other estimators of the recent SFR can be found in the literature (see Kennicutt 1998 for an interesting review on several methods to derive the SFR), we focus 
on only two of them, those using the NUV and FIR fluxes. 
Our aim is to compare commonly used recipes to derive SFR from  the UV and FIR luminosity of the galaxies. Therefore we will make very classical calculations, as described below. For consistency we re-derive the calibrations in a homogeneous way, adapted to the GALEX bands: the formulae are found to be very similar to those of Kennicutt (1998). 

The underlying physical justification for deriving the SFR of a galaxy from the UV luminosity is the following: most of the UV photons emerging from a galaxy originate in the atmospheres of stars younger than $\sim 10^{8}$~yr. Thus, the SFR is proportional to the UV luminosity emitted by the young stars under the assumption that the SRF is approximately constant over this timescale. This is reasonable given that Salim et al. (2005) and Burgarella et al. (2005) found that the intensity of the youngest burst in  large samples of nearby galaxies contributes typically less than 5\% to the total. 
However, the presence of dust absorbs a part of the UV light escaping from galaxies and breaks down the proportionality between the SFR and the observed UV luminosity. 

As star-forming galaxies may present a large variety of relative geometries between stars and dust, the scattering of the stellar photons through the interstellar medium may introduce a fraction of them in the line of sight before they escape the galaxy. Thus, the effect of the dust differs from a  pure extinction but is a complex   combination of absorption  and scattering. Following Gordon et al. (1997) we will use the term `dust attenuation' for this global process at work in galaxies. 
The most commonly accepted method to estimate the dust attenuation at UV wavelengths is to use the ratio of FIR-to-UV fluxes (Buat \& Xu 1996; Meurer et al. 1999; Gordon et al. 2000). Several analytical expressions are already available in the literature for different UV wavelengths (Panuzzo et al. 2003; Kong et al. 2004; Buat et al. 2005).
All these expressions are fairly consistent except at high values of the dust attenuation, where some dispersion appears (e.g. Meurer et al. 1999, Kong et al. 2004, Buat et al. 2005 at $\lambda \sim 1500$\AA.).
In this work we use the prescription of Buat et al. (2005) to obtain the corrected NUV and FUV luminosities:
\begin{equation}
A_{NUV} = -0.0495 x^{3} + 0.4718 x^{2} + 0.8998 x + 0.2269
\label{attenuv_eq}
\end{equation}
where $x = \log L_{IR}/L_{NUV}$ and
\begin{equation}
A_{FUV} = -0.0333 y^{3} + 0.3522 y^{2} + 1.1960 y + 0.4967
\label{attefuv_eq}
\end{equation}
where $y = \log L_{IR}/L_{FUV}$.

Once the observed NUV and FUV luminosities have been corrected for dust attenuation, the SFRs can be derived using the following expressions\footnote{This formula has been derived from Starburst99 (Leitherer et al. 1999) and assuming solar metallicity, and a Salpeter IMF from 0.1 to 100~$M_{\odot}$.}:

\begin{equation}
\log SFR_{NUV} (M_{\odot}~yr^{-1}) = \log L_{NUV,corr} (L_{\odot}) - 9.33
\label{sfrnuv_eq}
\end{equation}
\begin{equation}
\log SFR_{FUV} (M_{\odot}~yr^{-1}) = \log L_{FUV,corr} (L_{\odot}) - 9.51
\label{sfrfuv_eq}
\end{equation}

\begin{figure}
\epsscale{1.0}
\plotone{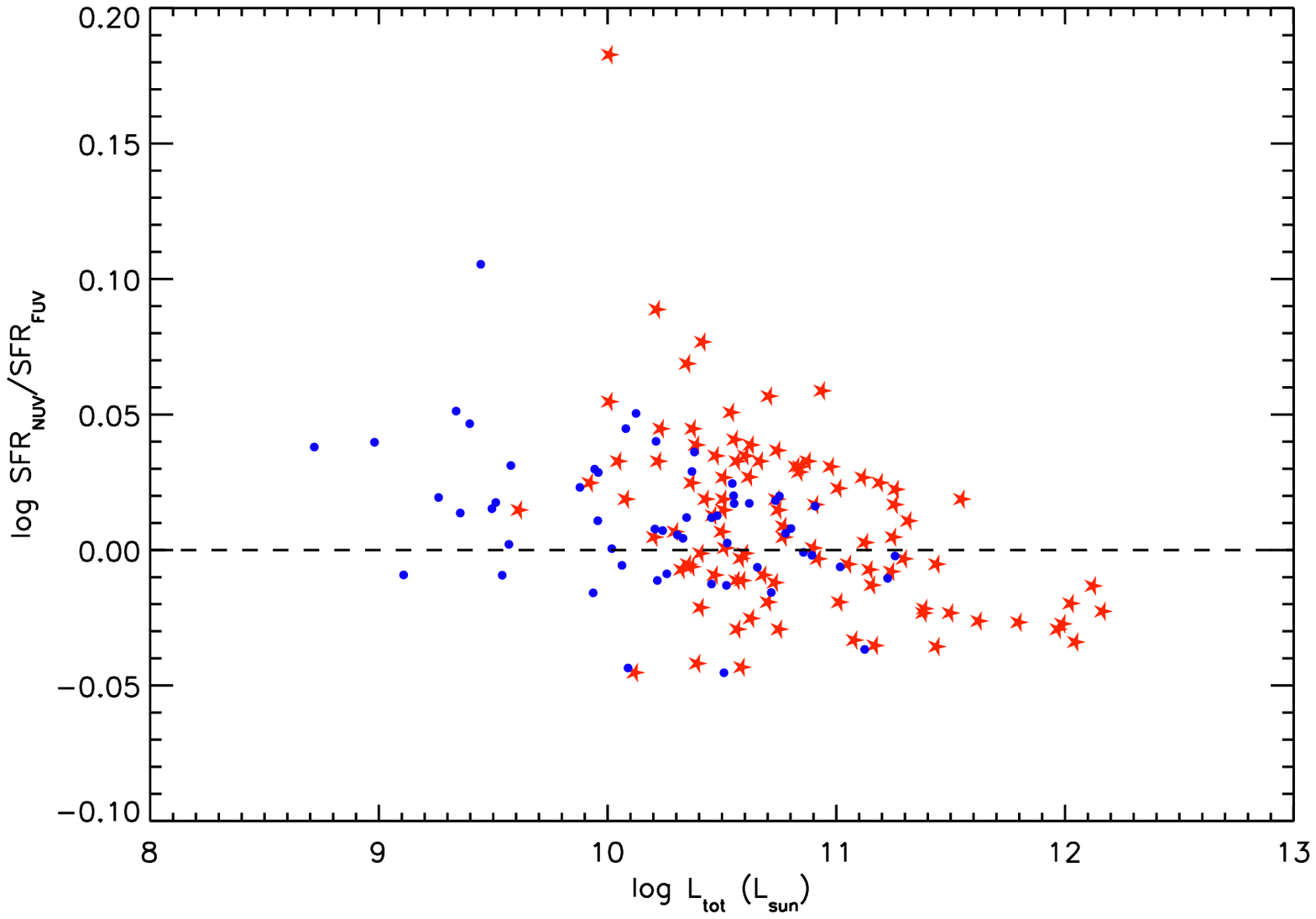}
\caption{$\log SFR_{NUV}/SFR_{FUV}$ vs. $L_{tot}$. 
Symbols are as in Figure~3.
\label{sfrnuvsfrnuvfuv}}
\end{figure}

In Figure~\ref{sfrnuvsfrnuvfuv} we show the ratio of $SFR_{NUV}/SFR_{FUV}$ as a function of $L_{tot}~(= L_{NUV} + L_{60})$,
which traces the bolometric luminosity related to recent star formation and has the advantage of being a purely observational quantity. As this figure shows, both quantities are equivalent with a dispersion of about 20\%. Since our sample is NUV-selected, hereafter we will use NUV as our reference wavelength for star formation related properties.

The luminosity at IR wavelengths provides a different avenue to the derivation of the SFR. 
Dust absorbs photons at UV wavelengths and re-emits most of them at IR wavelengths ($8 - 1000\mu$m). 
Under the hypothesis that all the UV photons are absorbed by dust, the IR luminosity would be a direct tracer of the SFR of a galaxy.
One source of uncertainty is the difficulty in estimating the total IR luminosity from the FIR flux at only one or two wavelengths. In this paper we use the prescription of Dale et al. (2001) and derive $L_{IR}$ by using $f_{60}$ and $f_{100}$. For the galaxies for which only $f_{60}$ is available we use the mean value of $f_{60}/f_{100}$ estimated using the galaxies detected at both wavelengths. 
If we assume the same scenario as for Eq.~\ref{sfrnuv_eq}, the SFR can be expressed as:
\begin{equation}
\log SFR_{dust} (M_{\odot}~yr^{-1}) = \log L_{IR} (L_{\odot}) - 9.75
\label{sfrfir_eq}
\end{equation}
However, Eq.~\ref{sfrfir_eq} is a good approximation only for the most extreme starbursts, since many of the FIR-selected galaxies are, in fact, detected at UV wavelengths. A further limitation of this method concerns the fraction of the total IR luminosity heated by old stars (the cirrus component, hereafter represented by $\eta$), which should be removed before applying Eq.~\ref{sfrfir_eq}. This quantity is known to depend on the morphological type of galaxies (Sauvage \& Thuan 1992), but a precise estimate for individual galaxies is subject to large uncertainties (Bell~2003).

\begin{figure}
\epsscale{1.0}
\plotone{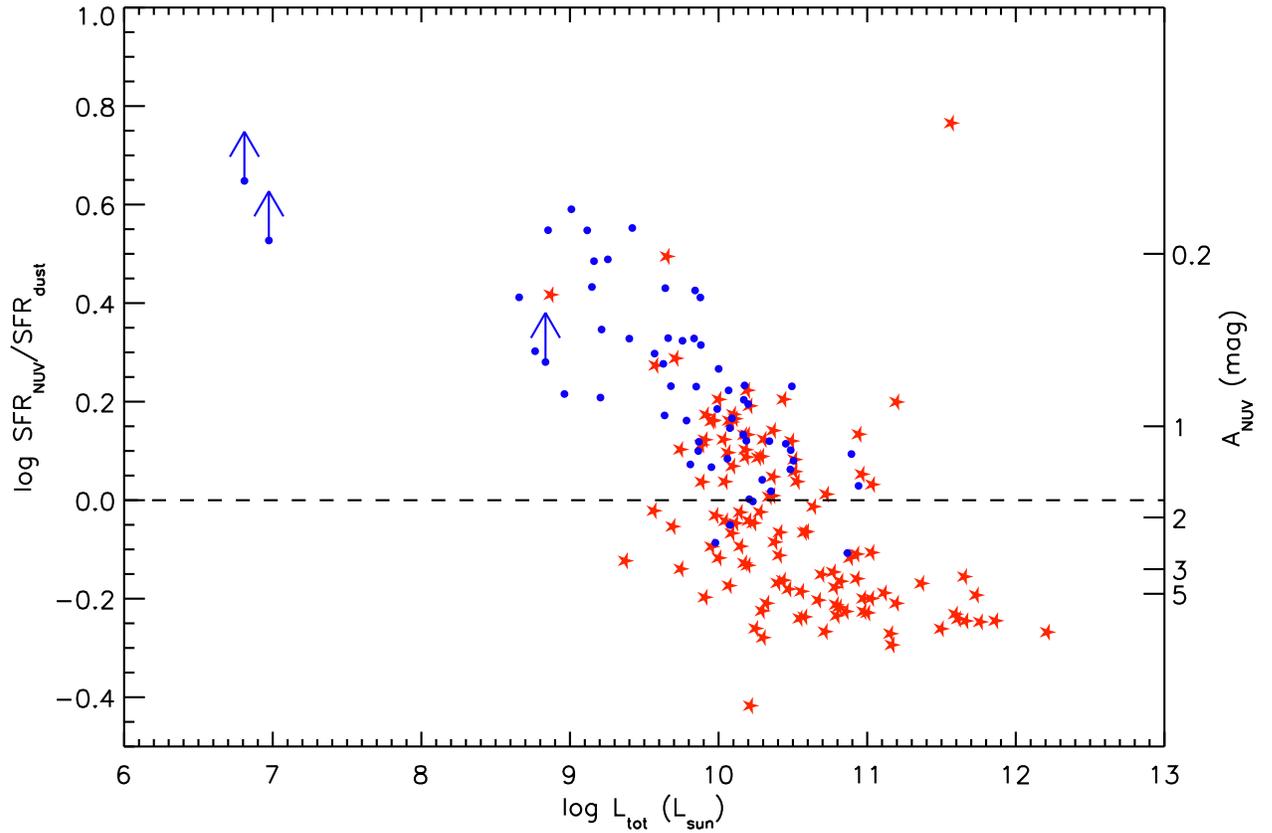}
\caption{$\log SFR_{NUV}/SFR_{dust}$ vs. $L_{tot}$. 
Symbols are as in Figure~3.
The right Y axis indicates the NUV dust attenuation.
\label{sfrnuvdustb}}
\end{figure}

The SFRs estimated from these methods are often compared in the
literature for individual objects or for large samples of galaxies. In
order to see whether they are consistent with each other we show here
a comparison of the two using the galaxies of our two
samples. Figure~\ref{sfrnuvdustb} shows the ratio
$SFR_{NUV}/SFR_{dust}$ as a function of $L_{tot}$; each sample shows a
different behavior. For the NUV-selected sample (blue, filled
circles), $SFR_{NUV}$ is always larger than $SFR_{dust}$ (and the ratio
can be as high as 3) but the discrepancy is lowered as $L_{tot}$ (and
$A_{NUV}$) increase. This result is expected since we have seen in
Section~3 that low luminous galaxies are brighter in the  NUV than at
60$\mu$m. This affirms that $SFR_{dust}$ cannot give a proper
estimation of the SFR for these galaxies.

The FIR-selected galaxies extend the trend found for the NUV-selected sample to higher luminosities.
For values of $L_{tot} \geq 3 \times 10^{10}$ (and for higher values of the dust attenuation), where no NUV-selected galaxies are present, $SFR_{dust}$ systematically exceeds $SFR_{NUV}$ by a factor of $\sim 2$.
One reason that could play a role in this inconsistency between the two estimators is that the dust attenuation is not properly estimated for very dusty galaxies. In any case it does not make sense to use the corrected UV luminosity to measure the SFR for these IR bright galaxies. In fact, Charmandaris et al. (2004) have reported decoupled IR and UV emissions for some dusty galaxies, which could be at the basis of the discrepancy found between $SFR_{NUV}$ and $SFR_{dust}$ found in this work for galaxies with large attenuation.

The conclusion of this analysis seems to be that $SFR_{NUV}$ is a good tracer of the SFR for low values of the attenuation, and in the opposite extreme $SFR_{dust}$ must be used for very heavily attenuated galaxies. There is no obvious way to delimit these two different regimes, or to chose which  and which of the two indicators should  be used in the intermediate cases. And so we warn users about any undiscriminated comparison of $SFR_{NUV}$ and $SFR_{dust}$ for samples of galaxies selected with different criteria. 

An alternative tracer of the SFR containing information from NUV and IR luminosities has already been discussed in the literature (Hirashita et al. 2003; Iglesias-P\'{a}ramo et al. 2004, Bell 2003):

\begin{equation}
SFR_{tot} = SFR^{0}_{NUV} + (1 - \eta) \times SFR_{dust}
\label{sfrtot_eq}
\end{equation}
where $\eta$ accounts for the IR cirrus emission and $SFR^{0}_{NUV}$ is obtained following Eq.~\ref{sfrnuv_eq} but using $L_{NUV,obs}$ (that is the observed NUV luminosity) instead of $L_{NUV,corr}$. 
This estimator has the advantages of being free of the model dependence of the attenuation correction, and it contains information of the observed NUV and the IR luminosities. 

One limitation of this estimator, $\eta$, is the adopted value of the
IR cirrus contribution.  Hirashita et al. (2003) and
Iglesias-P\'{a}ramo et al. (2004) reported a value of $\eta \sim 0.4$
for normal disk galaxies. Accurate values of $\eta$ for individual
galaxies are not easily obtained and instead, averaged values are often
used.  However, this parameter is strongly dependent on the sample of
galaxies under study and cannot be easily generalized. Whereas an
average value of $\eta \sim 0.4$ seems to apply for normal disk
galaxies, a value of $\eta \sim 0$ seems to better represent  the
properties of starbursts (Hirashita et al. 2003).  Bell (2003) also
proposed a cirrus correction for a compilation of galaxies from the
literature with FUV, optical, IR and radio luminosities. He found
$\eta \sim 0.32 \pm 0.16$ for galaxies with $L_{IR} \leq
10^{11}$~L$_{\odot}$ and $\eta \sim 0.09 \pm 0.05$ for galaxies with
$L_{IR} > 10^{11}$~L$_{\odot}$.  For our NUV-selected sample (similar
to the normal star forming galaxies of Hirashita et al.) a value of
$\eta \sim 0.2$ gives similar values for $SFR_{NUV}$ and
$SFR_{tot}(NUV)$. Although our NUV-selected sample must contain
galaxies more active than that of Hirashita et al. (since their
selection is based on optical fluxes rather than on UV fluxes), this
result gives an idea of the uncertainties related to the determination
of $\eta$.  For practical issues, throughout this paper we will use
the value of $\eta$ of Bell (2003) -- not far from that of Hirashita
et al. (2003) -- when computing $SFR_{tot}$, but keeping in mind that
the uncertainties reported by this author are of the order of 50\%.

\begin{figure}
\epsscale{1.0}
\plotone{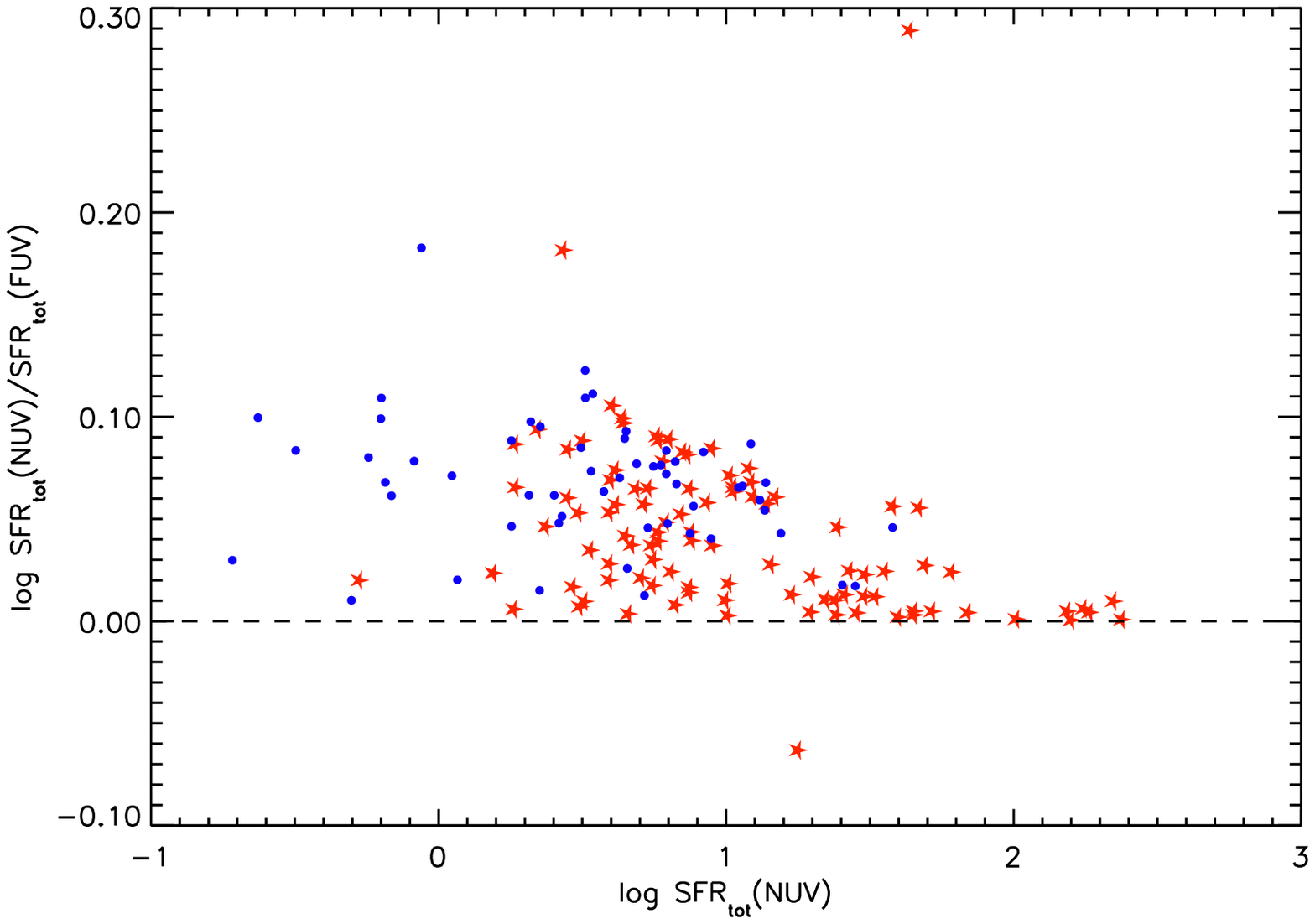}
\caption{$\log SFR_{tot}(NUV)/SFR_{tot}(FUV)$ vs. $\log SFR_{tot}(NUV)$. Symbols are as in Figure~3.
\label{ldustnuvldustfuv}}
\end{figure}

Another limitation of $SFR_{tot}$ is that it depends on the wavelength at which we measure the UV flux. In order to illustrate this point we show in Figure~\ref{ldustnuvldustfuv} the ratio of $SFR_{tot}(NUV)/SFR_{tot}(FUV)$ as a function of $SFR_{tot}(NUV)$ for both samples. 
As can be seen, for the NUV-selected galaxies $SFR_{tot}(NUV)$ is systematically larger than $SFR_{tot}(FUV)$ by about $20\%$. This 
discordance for the NUV-selected galaxies is due to the fact that the UV attenuation is not grey: $A_{NUV} \leq A_{FUV}$ for most galaxies (see Buat et al. (2005) and Table~\ref{photo}), and since we showed in Figure~\ref{sfrnuvsfrnuvfuv} that $SFR_{NUV} \approx SFR_{FUV}$, it is obvious that $SFR^{0}_{NUV} \geq SFR^{0}_{FUV}$.
On the contrary, for the brightest FIR-selected galaxies the agreement between $SFR_{tot}(NUV)$ and $SFR_{tot}(FUV)$ is good since for these galaxies $SFR_{tot}$ is dominated by $SFR_{dust}$. 
We conclude that  $SFR_{tot}$ is stable to within 20\% for whatever UV wavelength at which we measure the UV flux.

\begin{figure}
\epsscale{1.0}
\plotone{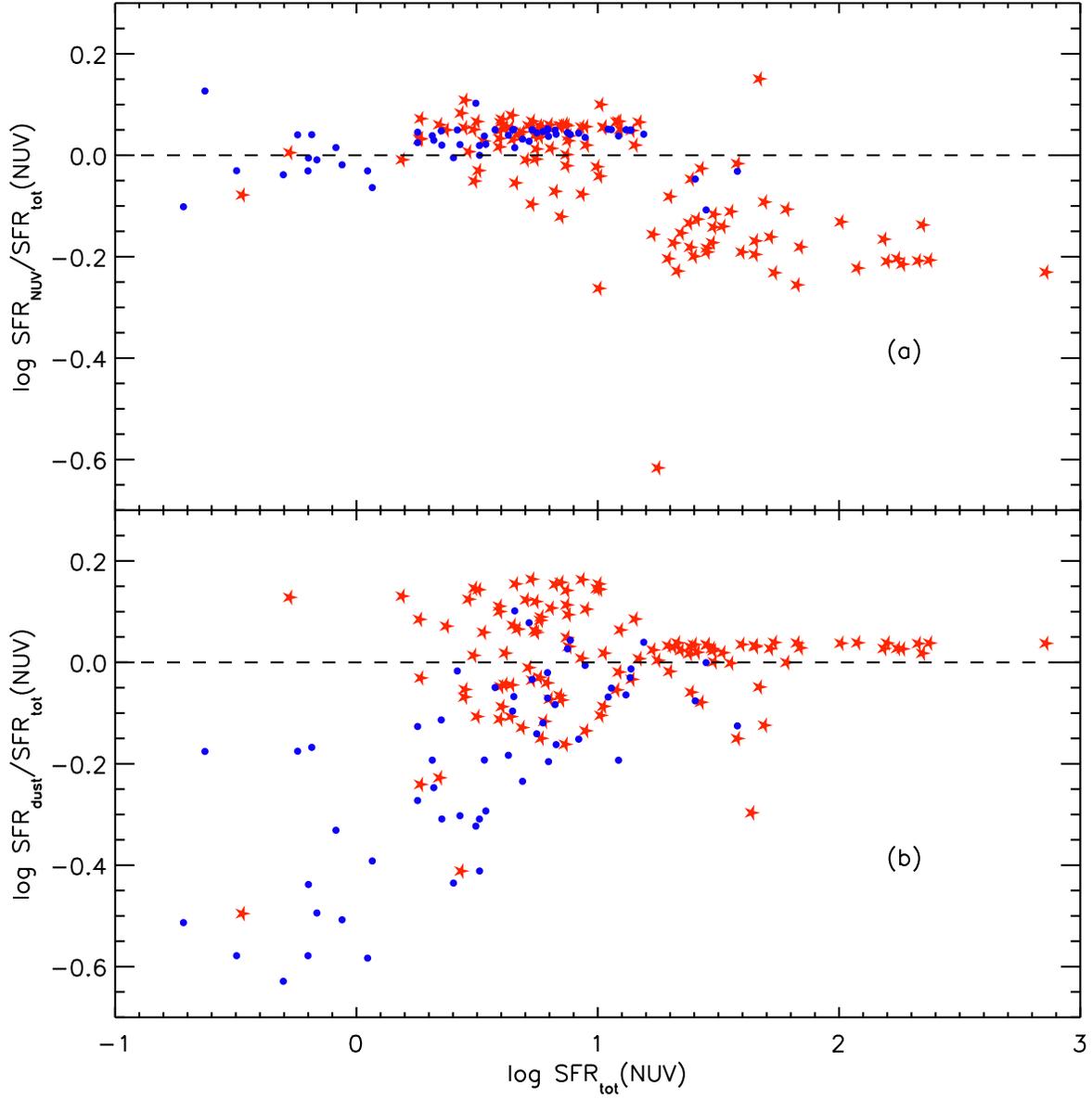}
\caption{(a) $\log SFR_{NUV}/SFR_{tot}(NUV)$ vs. $\log SFR_{tot}(NUV)$ (b) $\log SFR_{dust}/SFR_{tot}(NUV)$ vs. $\log SFR_{tot}(NUV)$. Symbols are as in Figure~3.
\label{sfrtotsfrnuv}}
\end{figure}

We compare now $SFR_{tot}(NUV)$ to the classical estimators $SFR_{NUV}$ and $SFR_{dust}$, in order to set their domain of applicability. 
Figure~\ref{sfrtotsfrnuv}a shows the comparison between $SFR_{NUV}$ and $SFR_{tot}$. At low values of the SFR both quantities are almost identical for the NUV-selected galaxies. This is expected since for these galaxies both $A_{NUV}$ and $L_{IR}$ are almost negligible and $SFR_{NUV} \approx SFR_{NUV}^{0}$. As the SFR grows, we note an increase of $SFR_{NUV}$ with respect to $SFR_{tot}$, but always within $\sim 15\%$. This increase could be due to the choice for the cirrus correction and/or to the fact that $A_{NUV}$ does not exactly corresponds to  the dust emission (since factors other than absorption do play a role in the attenuation, like for example the relative geometry between stars and dust). Finally, the NUV-selected galaxies with the largest values of SFR show a decrease of $SFR_{NUV}$ with respect to $SFR_{tot}$. We stress that these galaxies have $L_{IR} > 10^{11}$~L$_{\odot}$ and so their cirrus correction is different from for the rest. All in all we find that for the NUV-selected galaxies,  basically those with $SFR_{NUV} \leq 15$~$M_{\odot}$~yr$^{-1}$, $SFR_{NUV}$ and $SFR_{tot}$ are equivalent to within $\sim 15\%$.
The FIR-selected galaxies show a different behavior. Whereas those with $L_{IR} < 10^{11}$~L$_{\odot}$ show an $\sim 15\%$ excess of $SFR_{NUV}$ with respect to $SFR_{tot}$, similar to the NUV-selected galaxies, for those with $L_{IR} > 10^{11}$~L$_{\odot}$, $SFR_{NUV}$ is well below $SFR_{tot}$. This is easily understood as a consequence of the already mentioned discrepancy between $SFR_{dust}$ and $SFR_{NUV}$ for galaxies dominated by their IR emission.

In Figure~\ref{sfrtotsfrnuv}b we compare $SFR_{dust}$ and $SFR_{tot}$. The NUV-selected galaxies follow a very dispersed trend with $SFR_{dust}/SFR_{tot}$ increasing with $SFR_{tot}$. This behavior is due to the fact that $SFR_{dust}$ lacks the UV contribution which is dominant in these galaxies. The FIR-selected galaxies obey two different trends:  for galaxies with $SFR_{tot} \leq 15$~$M_{\odot}$~yr$^{-1}$ the ratio $\log SFR_{dust}/SFR_{tot} \sim 0$, although with a dispersion of $\sim 0.2$~dex. This large dispersion is due to the contribution of the NUV luminosity to $SFR_{tot}$, which is important for the less attenuated galaxies. On the contrary, at large values of $SFR_{tot}$ the average value of $\log SFR_{dust}/SFR_{tot} \sim 0.04$~dex with a very small dispersion. This is a consequence of the fact that most of these galaxies have $L_{IR} > 10^{11}$ and are dominated by their IR emission, so the difference between $SFR_{dust}$ and $SFR_{tot}$ corresponds basically to the cirrus correction applied to $SFR_{tot}$, which is minimal.

\begin{figure}
\epsscale{1.0}
\plotone{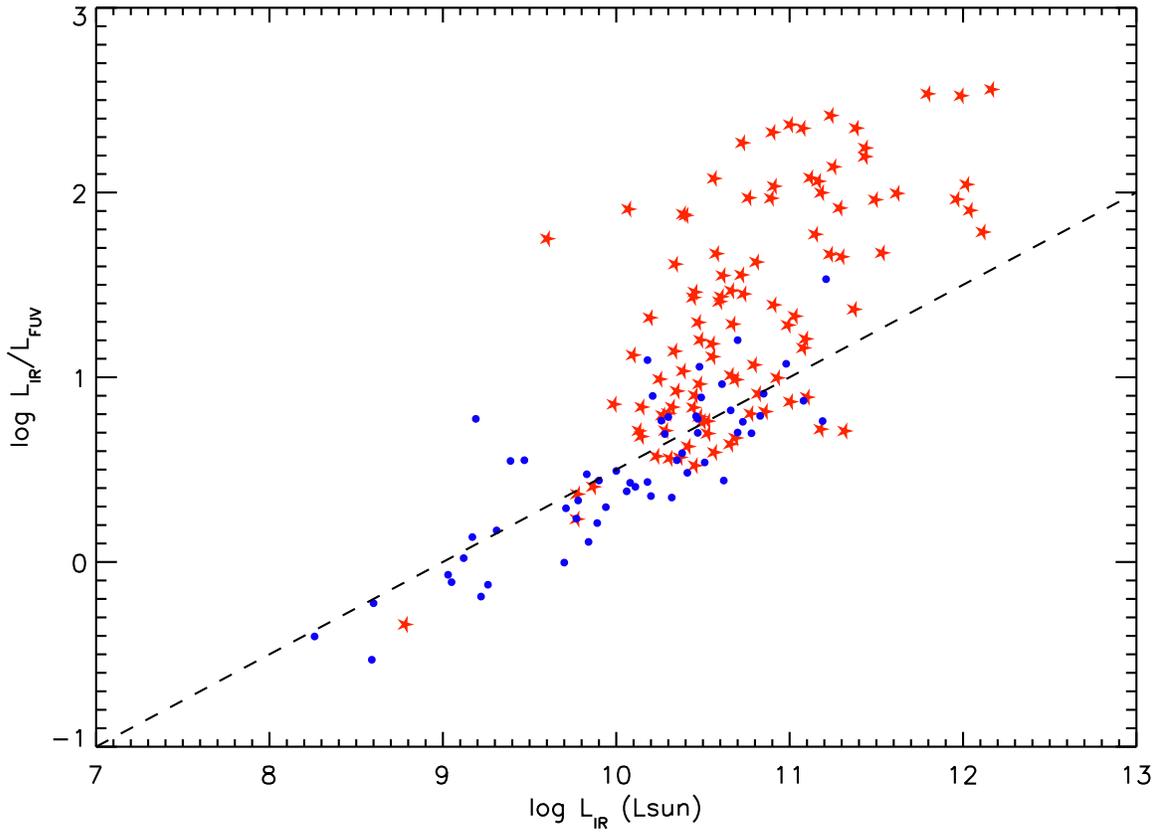}
\caption{$L_{IR}$ vs. $L_{IR}/L_{FUV}$. Symbols are as in Figure~3. The dashed line corresponds to the relation $L_{IR}/L_{FUV} \sim \sqrt{L_{IR}/10^9}$ of Bell (2003).
\label{lfuvtirltir}}
\end{figure}

Bell (2003) proposed a calibration of the SFR similar to the one described in Eq.~\ref{sfrtot_eq} but using FUV as the reference UV wavelength. His method is based on the relation he found between $L_{IR}/L_{FUV}$ and $L_{IR}$ ($L_{IR}/L_{FUV} \sim \sqrt{L_{IR}/10^9}$) 
for a compilation of galaxies from the literature with FUV, optical, IR and radio luminosities. 
One can see in  Figure~\ref{lfuvtirltir} that our NUV-selected galaxies follow well the Bell's relation whereas it is not the case for the FIR-selected sample. The galaxy sample used by Bell is therefore closer to a UV selection than to an IR one. Again we emphasize the importance of the selection biases in deriving SFRs.

The overall conclusion emerging from this study is that $SFR_{tot}$ seems to be a proper estimator of the SFR of galaxies whatever their dust content is, since it avoids the main problems of the clasical estimators $SFR_{NUV}$ and $SFR_{dust}$ and is consistent with them within their respective domains of applicability to within $\sim 15\%$. 
We again warn against indiscriminate comparisons of the SFR of galaxies estimated from these classical estimators since the results could be strongly affected by selection biases as we have illustrated in this section.
The combined uncertainty of $SFR_{tot}$ due to the choice of the UV wavelength at which we measure the UV flux and to the cirrus contribution to the IR luminosity is $\lesssim 55\%$. Throughout this paper, we adopt $SFR_{tot}(NUV)$ as our proxy to trace the recent SFR.

\subsection{Star formation history}

The determination of the SFR of a galaxy gives information about the total number of young stars that are being formed. But this does not necessarily mean that the light coming from this galaxy is dominated by these young stars given that  most galaxies are composed of a mixture of various stellar populations of different ages. 
This parameter is of major importance in understanding the SFH of the Universe. Recent results based on large amounts of SDSS data suggest that the higher the mass of a galaxy, the earlier its stars were formed (Heavens et al. 2004), thus supporting the so-called ``downsizing'' explanation for the SFH of galaxies already proposed by several authors (e.g. Cowie et al. 1996; Brinchmann \& Ellis 2000; Boselli et al. 2001). We devote this section to the study of the SFH of the galaxies in our samples. 

A quantitative estimation of the SFH of a galaxy requires information relating the relative contribution of young and old stars. 
The birthrate parameter (hereafter $b$) has been proposed as a quantitative estimator of the recent SFH of a galaxy (Scalo 1986). It is defined as the ratio between the current and the past-averaged SFR:
\begin{equation}
b = \frac{SFR}{\left< SFR \right>}
\label{b_eq}
\end{equation}

Since $b$ depends on the overall SFH of the galaxy, an accurate estimation 
from observational quantities is complex and involves several approximations. 
A detailed derivation of $b$ following the prescriptions of Boselli et al. (2001) can be found in Appendix~\ref{apendice_b}. As explained in Section~4.1, the NUV luminosity is sensitive to the SFR over a timescale of $\sim 10^{8}$~yr, and thus $b$ is not sensitive to shorter-timescale variations in the SFH. However, this is not a serious problem since Burgarella et al. (2005) have shown that less than 20\% of the galaxies in  either sample have bursts younger than $10^{8}$~yr.

\begin{figure}
\epsscale{1.00}
\plotone{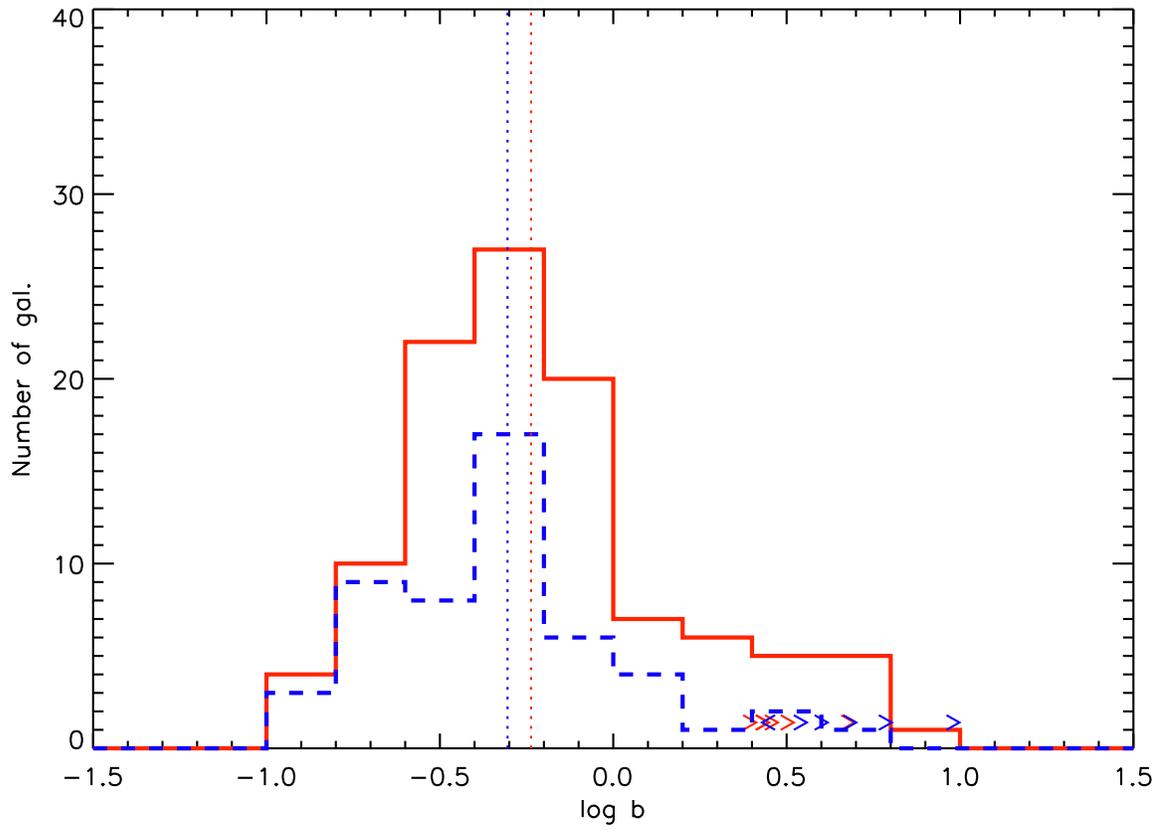}
\caption{Histogram of the distribution of $\log b$ for the NUV (blue dashed line) and the FIR (red solid line) selected samples. The vertical dotted lines correspond to the median values of the distributions.
Arrows correspond to nominal limiting values.
}
\label{histob}
\end{figure}

\begin{figure}
\epsscale{1.00}
\plotone{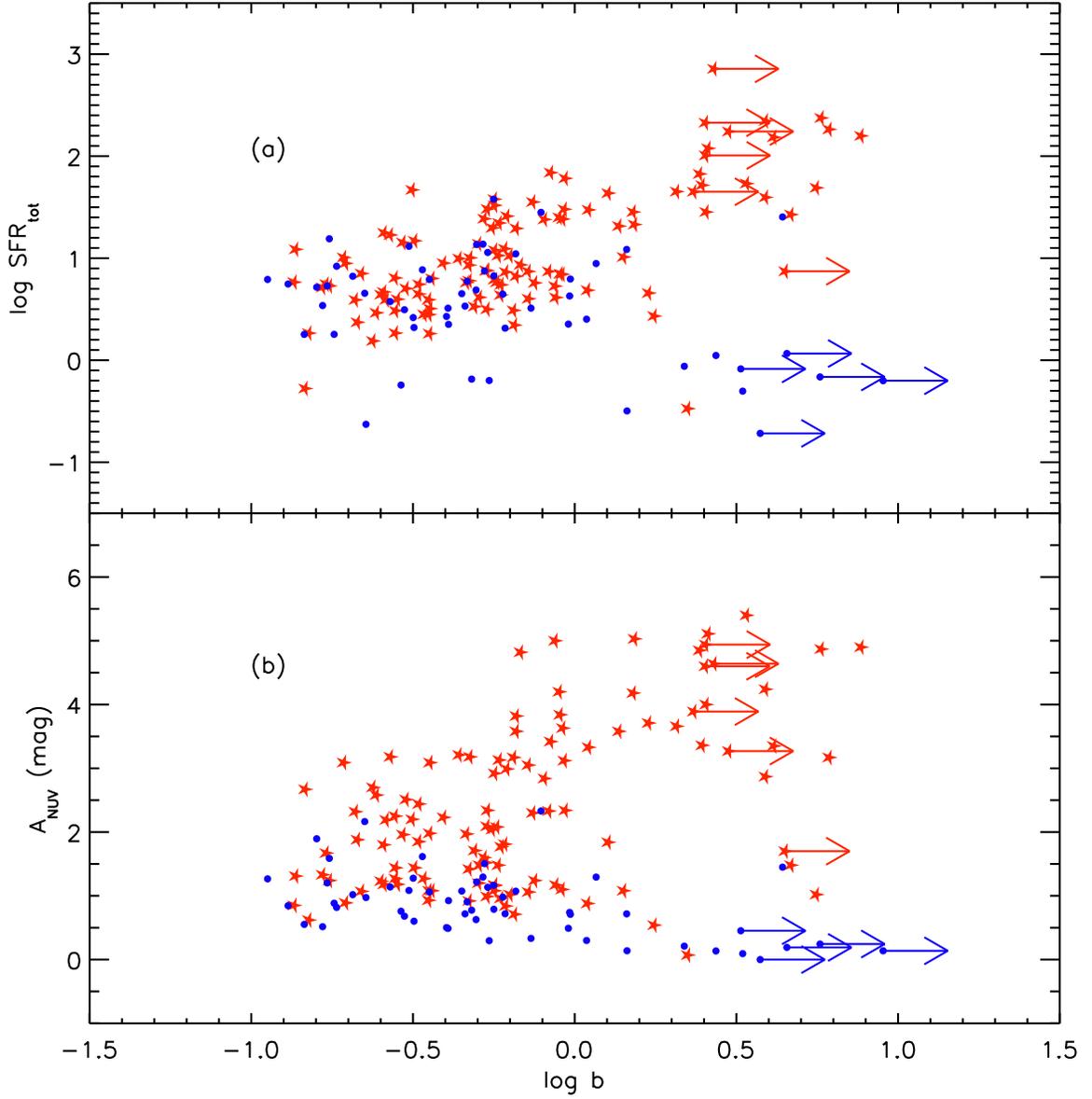}
\caption{(a) $SFR_{tot}$ vs. $b$. (b) $A_{NUV}$ vs. $b$. Symbols are as in Figure~3.
\label{bvssfrtot}}
\end{figure}

Figure~\ref{histob} shows the distributions of $b$ for both samples of galaxies. The median values of both distributions are similar: 0.50 and 0.58 for the NUV and FIR-selected galaxies respectively. 
In Figure~\ref{bvssfrtot}a we show the relation between $SFR_{tot}$ and $b$ for both samples. 
In the range of overlap between the two samples (approximately $0.5 \leq \log SFR_{tot} \leq 1.5$) the values of $b$ are consistent, but beyond this region two different trends are seen: the NUV-selected galaxies show no trend of b with the SFR, whereas the FIR-selected galaxies show an increase of $b$ for high SFR.
This bimodal behavior of $b$ is also seen in Figure~\ref{bvssfrtot}b, where $b$ is plotted as a function of the attenuation. Again, for the NUV-selected galaxies $b$ at lower values of the attenuation, although this trend is very dispersed.
The opposite holds for the FIR-selected galaxies, with galaxies with high $b$ being the most attenuated. 
Thus, the picture emerging from this study is that galaxies dominated by young stellar populations fall into two categories: those showing low SFRs and low attenuation, which naturally appear in UV surveys, and those with high SFRs and large attenuation, mainly detected in FIR surveys. 

\subsection{The link between the $H$ luminosity and star formation properties of galaxies}

The baryonic mass of galaxies is a key parameter in understanding their evolution. It has been proposed as the parameter which governs the SFH, rather than the morphological type, for example (Boselli et al. 2001). In addition, it is often used to derive some properties like the dust attenuation in semi-empirical models of formation and evolution of galaxies. For this reason we devote this section to a discussion of the effects of the sample selection on the relation between the mass and the star formation related properties of galaxies. As explained in the previous section, we will use the $H$-band luminosity as a tracer of the galaxy mass.

\begin{figure}
\epsscale{1.00}
\plotone{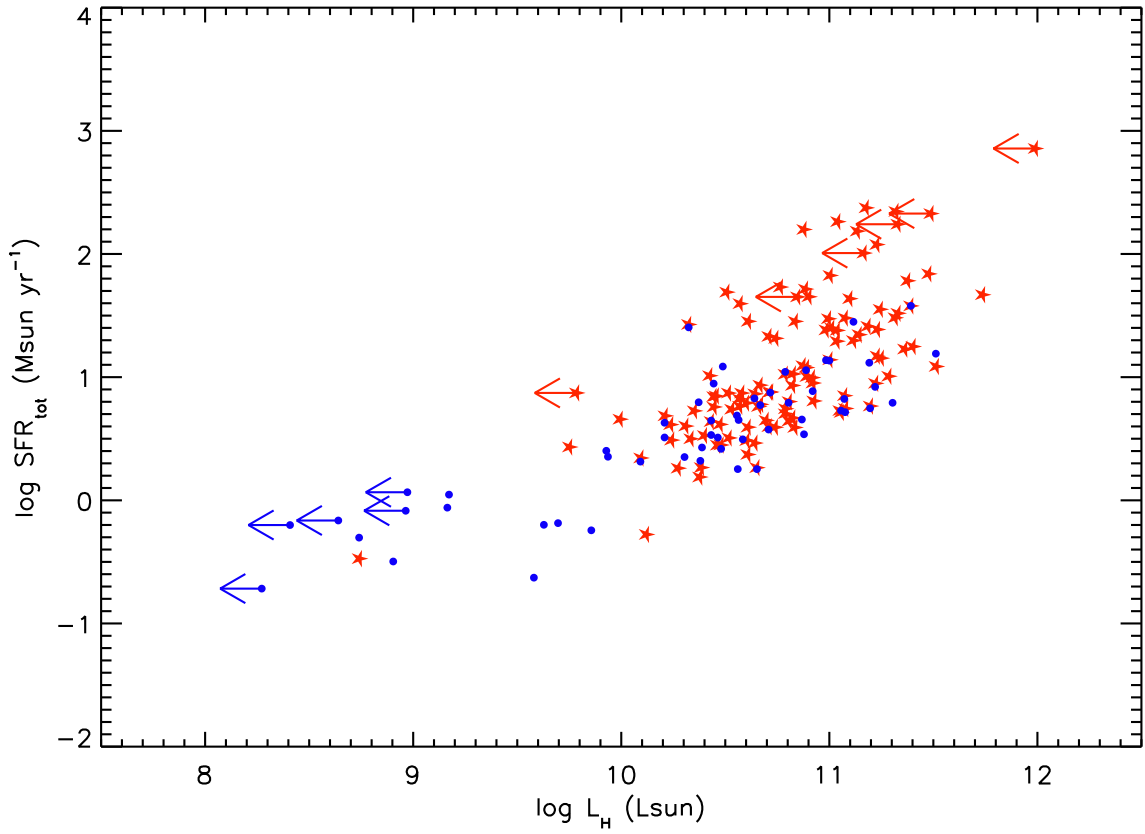}
\caption{$H$-band luminosity vs. $SFR_{tot}$. Symbols are as in Figure~3.
}
\label{lhsfrtot}
\end{figure}

First we show in Figure~\ref{lhsfrtot} $SFR_{tot}$ as a function of the $H$-band luminosity. Both samples show a positive relation between these two quantities, which means that more massive galaxies are also currently forming more young stars. 
This result is expected since we are comparing two extensive quantities. 
However, whereas the relation followed by the NUV-selected galaxies  shows a small dispersion, the FIR-selected galaxies exhibit a more dispersed relation, especially at the most massive end. At high galaxian masses the range in SFR spans almost two orders of magnitude, which is not seen in the NUV-selected sample. 

\begin{figure}
\epsscale{1.00}
\plotone{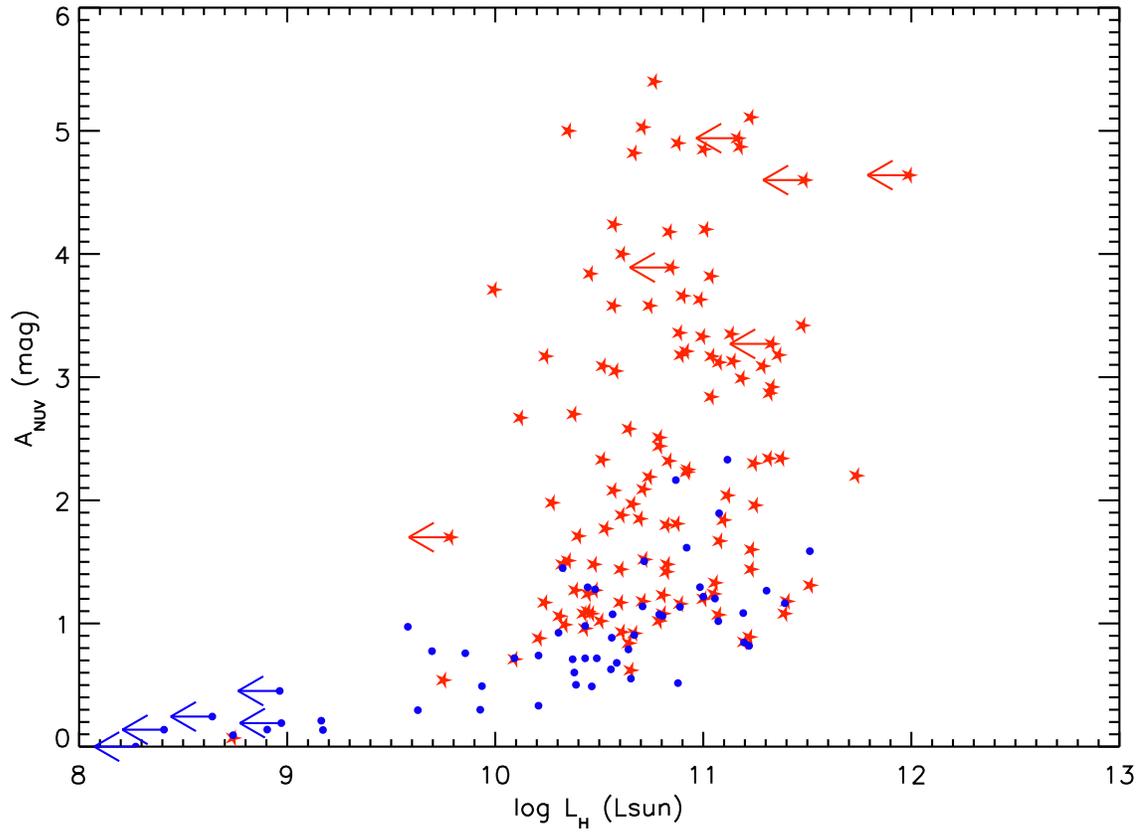}
\caption{$H$-band luminosity vs. $A_{NUV}$. Symbols are as in Figure~3.
}
\label{hvsanuv}
\end{figure}

In Figure~\ref{hvsanuv} we show the dust attenuation as a function of the $H$-band luminosity. Two different trends are seen in this plot. 
The NUV-selected galaxies show a fairly good correlation between the two quantities, with the dispersion increasing towards high $H$-band luminosities. On the contrary, the FIR-selected galaxies span an interval of almost 5~mag in dust attenuation and no correlation at all is shown with the galaxian mass. While the trend followed by the NUV-selected galaxies could be interpreted as a result of the mass -- metallicity relation reported for samples of spiral and irregular galaxies (Garnet \& Shields 1987; Zaritsky 1993) in the sense that more metallic galaxies contain more dust, there is no simple explanation for the lack of any trend shown by the FIR-selected galaxies.

\begin{figure}
\epsscale{1.00}
\plotone{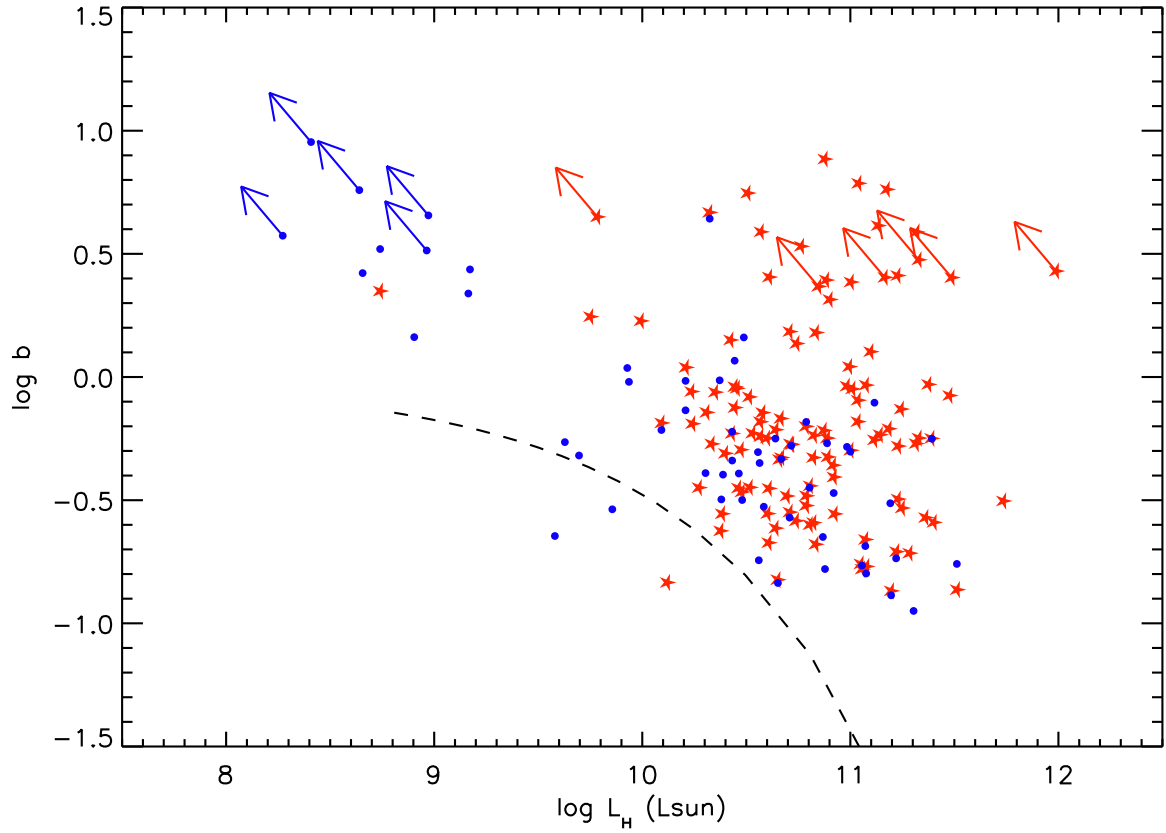}
\caption{$H$-band luminosity vs. $b$. Symbols are as in Figure~3. The dashed line corresponds to the calibration of Boselli et al. (2001) for their sample of late-type galaxies from nearby clusters with a normal H{\sc i} content.
}
\label{lh_b}
\end{figure}

Finally, we show in Figure~\ref{lh_b} the $b$ parameter as a function of the $H$-band luminosity. The NUV-selected galaxies follow the classical trend that low-luminosity galaxies have larger values of $b$ (e.g. Boselli et al. 2001). Some of the FIR-selected galaxies also follow this trend, although 
about 20\% of them that show large masses and large values of $b$. As shown in Figure~\ref{hvsanuv}, these galaxies are among the most attenuated of the FIR-selected sample.
Overall, our galaxies are shifted towards higher values of $b$ with respect to the sample of galaxies of Boselli et al. (2001). These authors adopt a slightly different IMF than we do ($M_{up} = 80$~M$_{\odot}$ against 100~M$_{\odot}$) and different evolutionary synthesis codes. Nevertheless, the large  shift in $b$  found between the samples can probably not  be explained by these differences alone.
The correction for dust attenuation could also partially explain the shift in $b$, since Boselli et al. assume average values of 0.20~mag for Sds and later types, and  0.60~mag for types earlier than Sd. For our NUV and FIR-selected samples, the values of the dust attenuation estimated from the FIR/UV flux ratio of each of the individual galaxies show higher averaged values for the two categories of morphologies than those of Boselli et al., which would imply higher SFRs. However, this effect is diluted by the fact that for many objects in their sample, Boselli et al. estimate the SFR as the mean value of $SFR_{\rm H \alpha}$ and $SFR_{UV}$.
A further factor that could be responsible for the shift in $b$ is the different selection effect of the sample: the sample of Boselli et al. (2001) is drawn from the nearby clusters Virgo, Cancer, Coma and A1367 and from the Coma-A1367 supercluster. Although not a unique selection criterion was applied, this sample can be defined as an optically selected sample of galaxies with a normal H{\sc i} content. Thus, in their sample there is a non-negligible fraction of Sa-Sab, bulge-dominated galaxies, which tend to lower the average value of $b$ (see their Figure~2). Since our selections are based on NUV and FIR fluxes, we argue that we are surely avoiding these kind of objects. Anyway, one important point is that  the NUV-selected galaxies follow the same relation between mass and $b$ as the optically selected ones (disregarding the absolute calibration of both quantities) and that a fraction of the FIR-selected galaxies do not follow this trend.

We have seen that the relation of the star-formation-related properties with the mass of galaxies strongly depends on the selection procedure of the sample: whereas for NUV-selected galaxies low-luminosity galaxies are also low mass, show low attenuation and have high values of $b$. A selection based on the FIR fluxes yields a different result: a population having high attenuation, high mass and strong star-formation activity appears. This population is absent in the NUV-selected sample. Since these galaxies present very high values of the attenuation (most of them are LIRGs and/or ULIRGs), their UV (and optical) fluxes are strongly dimmed and for this reason they are often excluded from flux limited surveys. However, even if these galaxies show such extreme properties, they do not put into question the downsizing picture for the SFH of galaxies since their contribution to the local cosmic SFR density is very low (see Takeuchi et al. 2005).

\section{The local cosmic SFR density from different estimators}

We saw in Section~4.1 that a proper estimation of the SFR is not possible with information restricted only to either NUV or FIR fluxes. However, big surveys usually provide large amounts of data only at single wavelengths, thus an estimation of the density of SFR over cosmological volumes has to be carried out under these constraints. The accuracy of the cosmic density of SFR has already been discussed by Hirashita et al. (2003) using FOCA UV data. In this section we make a similar analysis using the new GALEX data.

The usual way to estimate the average SFR density is to construct the monochromatic LF and then to weight the corresponding contribution to the SFR at a given luminosity with the probability of finding a galaxy with this luminosity. 

\begin{equation}
\rho_{\lambda} = \kappa_{\lambda} \int L_{\lambda} \phi(L_{\lambda}) dL_{\lambda}
\label{sfrdensity}
\end{equation}
where $\rho_{\lambda}$ is the SFR density estimated from the flux at a given $\lambda$, $\kappa_{\lambda}$ is the conversion factor between $SFR_{\lambda}$ and $L_{\lambda}$, and $\phi(L_{\lambda})$ is the LF.

A simple calculation using Eqs.~\ref{sfrnuv_eq}, \ref{sfrdensity} and the NUV LF of Wyder et al. (2005) yields a value of $\rho_{NUV} = 0.009^{+0.007}_{-0.004}$~M$_{\odot}$~yr$^{-1}$~Mpc$^{-3}$ for the local cosmic SFR density. Unfortunately, there is not a simple relation linking the observed NUV luminosity and the attenuation (see Figure~3 in Buat et al. 2005), so we adopt the median attenuation of our NUV-selected sample, which is $A_{NUV} = 0.78$~mag. 
After correcting for this median attenuation we obtain $\rho_{NUV,corr} = 0.018^{+0.013}_{-0.008}$~M$_{\odot}$~yr$^{-1}$~Mpc$^{-3}$.

Takeuchi et al. (2005) report a value of the local cosmic star formation density derived from $L_{IR}$ of $\rho_{IR} = 0.013^{+0.008}_{-0.005}$~M$_{\odot}$~yr$^{-1}$~Mpc$^{-3}$. However, we recall that this quantity does not account for the fraction of UV escaping photons or for the cirrus IR heating. For our FIR-selected sample, the median contribution of the NUV escaping luminosity to $SFR_{tot}$ is 17\%
which leads to $\rho'_{IR} = 0.016^{+0.010}_{-0.006}$~yr$^{-1}$~Mpc$^{-3}$.
After correcting for  the cirrus contribution assuming $\eta = 0.32$ we obtain $\rho_{IR,corr} = 0.011^{+0.007}_{-0.004}$~M$_{\odot}$~yr$^{-1}$~Mpc$^{-3}$ which is well below $\rho_{NUV,corr}$, although still within the 1-$\sigma$ uncertainty. Although we have previously accepted that a cirrus correction of $\eta = 0.32$ is valid for galaxies with $L_{IR} \leq 10^{11}$~L$_{\odot}$ and $\eta = 0.09$ for brighter galaxies, we argue that adopting just $\eta = 0.32$ for all the galaxies is not a bad approximation for this calculation since it can be seen in Takeuchi et al. (2005) that the contribution of $L_{IR} \phi(L_{IR})$ to the total $\int L_{IR} \phi(L_{IR}) dL_{IR}$ of galaxies with $L_{IR} > 10^{11}$~L$_{\odot}$ is very low and can hardly affect our calculations.

Proceeding in an analogous way as in Section~4.1, we can add both contributions to get the total cosmic SFR, and we get $\rho_{tot} = \rho_{NUV} + (1 - \eta) \times \rho_{IR}~(= 0.009 + 0.009) = 0.018$~M$_{\odot}$~yr$^{-1}$~Mpc$^{-3}$, which is  in good agreement with $\rho_{NUV,corr}$, which means that correcting $\rho_{NUV}$ with a median attenuation is a good approximation. 
We also stress that the $\rho_{tot}$ is almost equally shared between the NUV and the IR contribution.

The discrepancy found with $\rho_{IR}$ alone may be due to the average corrections applied. In order to obtain a better agreement between $\rho_{NUV,corr}$, $\rho_{tot}$ and $\rho_{IR,corr}$ two points should be studied in more detail:
\begin{itemize}
\item A detailed knowledge of the cirrus component is required since this contribution is probably multivalued for a given value of $L_{IR}$. Although the morphological type seems to drive this parameter (Sauvage \& Thuan 1992), it could also be dependent on $b$ since this parameter also measures the relative weight of the young and old stellar populations.
A more detailed study of a large samples of galaxies, covering a wide range of galaxian properties, could shed light on its fractional contribution to the total cosmic $L_{IR}$ density.
\item The bivariate LF $\phi(L_{NUV},L_{IR})$ appears to be the best way to estimate the fraction of UV photons escaping from the galaxy, required to correct $\rho_{IR}$. It is also required since for large values of $L_{NUV}$, the attenuation can take on multiple values, and thus an average value, as the one used in this work, might be not the most appropriate.
\end{itemize}
Under these conditions, the cosmic SFR densities would be expressed as:
\begin{equation}
\rho_{NUV,corr} = \int\int \kappa_{NUV} \times L_{NUV} \times 10^{A_{NUV}(L_{NUV},L_{IR})/2.5} \times \phi(L_{NUV},L_{IR}) dL_{NUV} dL_{IR}
\end{equation}
and
\begin{equation}
\rho_{IR,corr} = \int\int \kappa_{IR} \times \left[1 - \eta(L_{IR},L_{NUV})\right] \times L_{IR} \times (1 + (\kappa_{NUV}/\kappa_{IR}) \times (L_{NUV}/L_{IR})) \times \phi(L_{NUV},L_{IR}) dL_{NUV} dL_{IR}
\end{equation}
or if we use the approximation of Eq.~\ref{sfrtot_eq}
\begin{equation}
\rho_{tot} = \int\int \left( \kappa_{IR} \times \left[1 - \eta(L_{IR},L_{NUV})\right] \times L_{IR} + \kappa_{NUV} \times L_{NUV} \right) \times \phi(L_{NUV},L_{IR}) dL_{NUV} dL_{IR}
\end{equation}

\section{Conclusions}

We have performed a detailed study of the star formation properties of two samples of galaxies selected on the basis of their NUV and FIR fluxes, which were found to be representative of the nearby Universe when compared to samples drawn from larger volumes.
The main conclusions of this work are:
\begin{enumerate}
\item $L_{NUV}$ and $L_{60}$ are tightly correlated for the NUV-selected galaxies. The opposite holds for FIR-selected galaxies, which span a large range of $L_{60}$ for a given value of $L_{NUV}$ and show larger values of attenuation. Intrinsically bright galaxies are more luminous at FIR than at NUV wavelengths, including the UV luminous galaxies (UVLGs), and they show moderate to high attenuation.
\item The SFR deduced from the NUV fluxes, corrected for the dust attenuation ($SFR_{NUV}$), are not found to be consistent with those calculated using  the total dust emission ($SFR_{dust}$). Whereas $SFR_{NUV}$ is larger than $SFR_{dust}$ for galaxies with low attenuation ($A_{NUV} \lesssim 1$~mag) the inverse is found for bright, but highly extinguished galaxies, mostly selected in IR: $SFR_{NUV}$ is likely to underestimate the actual SFR in these galaxies by a factor $\sim 2$.
A combined estimator based on UV and IR luminosities with a cirrus correction depending on the IR luminosity seems to be the best proxy over the whole range of values of SFR. As a practical recipe we found that $SFR_{tot}$ and $SFR_{NUV}$ yield similar results for $SFR_{tot} \lesssim 15$~M$_{\odot}$~yr$^{-1}$, whereas $SFR_{tot}$ and $SFR_{dust}$ are almost equivalent for $SFR_{tot} \gtrsim 15$~M$_{\odot}$~yr$^{-1}$. 
\item NUV-selected galaxies follow a trend whereby low-mass galaxies show lower SFRs, low attenuation and higher values of $b$, indicating the existence of a dominant young stellar population. On the contrary, 
about 20\% of
the FIR-selected sample shows  high attenuation, high SFRs and also large values of $b$, most of them being LIRGs and/or ULIRGs. In spite of their discordant properties, these galaxies are not sufficiently abundant in the local Universe to question the downsizing picture for the SFH seen at $z = 0$ from optical surveys.
\item The cosmic SFR densities of the local Universe, estimated from the NUV and IR luminosities, are consistent to within 1-$\sigma$, although the difference between the two values is large, when average corrections for the attenuation, UV escaping photons and IR cirrus component are applied. The sum of the individual contributions is quite consistent with the value obtained from the NUV luminosities corrected for average attenuation.
A better knowledge of the cirrus contribution to $L_{IR}$ and of the bivariate LF is required in order to better understand the large differences found between the monochromatic estimators of the local SFR density.
\end{enumerate}

\acknowledgments

GALEX is a NASA Small Explorer, launched in 2003 April. We gratefully acknowledge NASA's support for construction, operation, and science analysis for the GALEX mission, developed in cooperation with the Centre National d'Etudes Spatiales of France and the Korean Ministry of Science and Technology.
This publication makes use of data products from the Two Micron All Sky Survey,
which is a joint project of the University of Massachusetts and the Infrared 
Processing and Analysis Center/California Institute of Technology, 
funded by the National Aeronautics and Space Administration and the National 
Science Foundation. This research has made use of the NASA/IPAC Extragalactic Database (NED) which is operated by the Jet Propulsion Laboratory, California Institute of Technology, under contract with the National Aeronautics and Space Administration.
The Lyon Extragalactic Database (LEDA) is available at http://leda.univ-lyon1.fr/.
TTT has been financially supported by the Japan Society for the Promotion of
Science.

\appendix

\section{Relation of the Mean Redshift of Galaxies and the LF: 
Representativity of a Survey Depth\label{depth}}

We show the strong dependence of the mean redshift on the shape of
the LF.
This means that the mean distance of galaxies does not represent the 
depth of a survey, but rather reflects an intrinsic property of the sample.

Since we treat a local sample of galaxies, we first approximate their 
distance by the classical Hubble's law:
\begin{eqnarray}
  cz \simeq \frac{r}{H_0} \;,
\end{eqnarray}
where $c$ is the velocity of light, and $r$ represents the distance.
We define ${\cal N}$ to be  surface density of galaxies on the sky, and
denote the LF as an explicit function of the characteristic luminosity, $L_*$
as $\phi(L/L_*)$.
Then, ${\cal N}$ is written as (Peebles 1993):
\begin{eqnarray}
  {\cal N} =  \int \int \phi \left[ \frac{L(r)}{L_*} \right] 
    \frac{dL(r)}{L_*} r^2 dr \;.
\end{eqnarray}
Using the detected flux density $S$, this can be expressed as
\begin{eqnarray}
  {\cal N} &=&  \int \int \phi \left[ \frac{4\pi}{L_*} 
    \left( \frac{c}{H_0}\right)^2 z^2 S \right] 
    \frac{4\pi}{L_*} \left( \frac{c}{H_0}\right)^2 z^2 dS 
    \left( \frac{c}{H_0}\right)^3 z^2 dz \nonumber \\
  &=&  \int \int \phi \left[ \frac{4\pi}{L_*} 
    \left( \frac{c}{H_0}\right)^2 z^2 S \right]
    \frac{4\pi}{L_*} \left( \frac{c}{H_0}\right)^5 z^4 dS dz \;.
\end{eqnarray}
We observe the distribution function of the sources 
with a {\sl fixed} flux density $S$ as
\begin{eqnarray}\label{eq:dif_nc}
  \frac{d^2{\cal N}}{dzdS} =  \phi \left( \alpha  z^2 S \right)
   \alpha \left( \frac{c}{H_0}\right)^3 z^4 \;,
\end{eqnarray}
where
\begin{eqnarray}
  \alpha \equiv \frac{4\pi}{L_*}\left(\frac{c}{H_0}\right)^2\;,
\end{eqnarray}

The mean redshift of a flux-limited survey (limiting flux density $S$),
$\langle z \rangle_{>S}$, is then
defined as\footnote{Note that this is a different quantity defined by
Equation~(5.134) of Peebles (1993).}
\begin{eqnarray}\label{eq:meanz}
  \langle z \rangle_{>S} \equiv \frac{\displaystyle \int_S^\infty \int_0^\infty
    \frac{d^2{\cal N}}{dzdS'} zdzdS'}{\displaystyle \int_S^\infty \int_0^\infty
    \frac{d^2{\cal N}}{dzdS'} dzdS'} \;.
\end{eqnarray}
The numerator is obtained as
\begin{eqnarray}\label{eq:numer}
  \int_S^\infty \int \frac{d^2{\cal N}}{dzdS'} zdzdS' &=& 
    \frac{1}{2\alpha^2} \left(\frac{c}{H_0}\right)^3
    \left[ \int_0^{\infty} \phi (x) x^2 dx\right] 
    \int_S^\infty {S'}^{-3} dS'\nonumber \\
  &=& \frac{1}{2\alpha^2} \left(\frac{c}{H_0}\right)^3
    \left[ \int_0^{\infty} \phi (x) x^2 dx\right] \frac{S^{-2}}{2}\;.
\end{eqnarray}
Here, $x \equiv \alpha z^2 S'$, a luminosity normalized by the characteristic 
luminosity of the LF, $L_*$ and expressed in terms of flux density $S'$.
Since the luminosity must be positive, the lower bound on the
integration with respect to $x$ is 0, and upper bound is 
 large, effectively taken to be $+\infty$.
Similarly the denominator becomes
\begin{eqnarray}\label{eq:denom}
  \int_S^\infty \int_0^\infty \frac{d^2{\cal N}}{dzdS'} dzdS' &=& 
    \frac{1}{2\alpha^{3/2}} \left(\frac{c}{H_0}\right)^3
    \left[ \int_0^\infty \phi (x) x^{3/2} dx\right] 
    \int_S^\infty {S'}^{-5/2} dS' \nonumber \\
  &=& \frac{1}{2\alpha^2} \left(\frac{c}{H_0}\right)^3
    \left[ \int_0^\infty \phi (x) x^2 dx\right] \frac{2S^{-3/2}}{3}\;.
\end{eqnarray}
Both the numerator and the denominator of this part is the moment of the LF.
This means that this is a function of its shape.

Combining Equations~(\ref{eq:meanz}), (\ref{eq:numer}), and (\ref{eq:denom}), 
we have
\begin{eqnarray}\label{eq:meanz2}
  \langle z \rangle_{>S} &=& 3\alpha^{-1/2}
    \frac{\displaystyle \int_0^\infty \phi (x) x^2 dx}{
    \displaystyle \int_0^\infty \phi (x) x^{3/2} dx} S^{-1/2} \nonumber \\
  &=& 3 \left(\frac{L_*}{4\pi}\right)^{1/2}\left(\frac{H_0}{c}\right)
    \frac{\displaystyle \int_0^\infty \phi (x) x^2 dx}{
    \displaystyle \int_0^\infty \phi (x) x^{3/2} dx} S^{-1/2} \;.
\end{eqnarray}

Let us carefully examine Equation~(\ref{eq:meanz2}).
First, the dependence of the mean redshift on the limiting flux density $S$
is a power of $-1/2$.
Also, it has the same order of dependence on $L_*$.
In contrast, the integral part of Equation~(\ref{eq:meanz2}) has an important
meaning.
Since this part depends on the second-order moment, 
the tail of the LF affects the value very strongly.
As we mentioned, we integrate over the (normalized) luminosity up to a
certain very large value, and this part is a ratio between the moments of 
order $3/2$ and 2.
Hence the contribution from a large value of $x$ controls the value.
Consequently, the mean redshift $\langle z \rangle_{>S}$ is very 
sensitive to the LF shape.

This aspect is clearly seen in the comparison of the expected redshift 
distributions in NUV and $60\mu$m calculated from the LFs and the limiting
flux density or magnitude, because the shapes of the LFs at these wavelengths
are very different (Buat \& Burgarella 1998; Takeuchi et al. 2005).
We show the comparison in Figures~\ref{histo_velo_all}b and ~\ref{histo_velo_all}c.
In these figures we fix the limiting flux density at $60\mu$m to 
be 0.6~Jy, while we change the limiting magnitude  from $AB_{NUV}=16.0$~mag
(the actual value we adopt in this work) to 18.0~mag.
The medians for the two wavelengths are very different when we adopt a limiting value of $AB_{NUV}=16.0$~mag. Only when we adopt a limiting value of $AB_{NUV}=18.0$~mag, which corresponds to a very sensitive survey, the median values approach for the NUV and FIR-selected samples.

\section{Estimating the birthrate parameter \label{apendice_b}}

Boselli et al. (2001) give a detailed recipe to estimate $\left< SFR \right>$, based only on observable quantities and/or adopted values for the parameters (see Gavazzi et al. 1996). Here we follow their prescriptions, using the $H$-band luminosity to estimate the past averaged SFR and the same parameters as Boselli et al. (2001):
\begin{equation}
b = \frac{SFR \times t_{0} \times (1 - R)}{L_{H} \times (M_{tot}/L_{H}) \times DM_{cont}}
\end{equation}
where $SFR$ in this work is averaged over $10^{8}$~yr, $t_{0}$ is the age of the disk (assumed to be equal to 13~Gyr), $R$ is the fraction of gas re-injected by stars through stellar winds into the interstellar medium during their lifetime (taken to be equal to 0.3 for a Salpeter IMF), $L_{H}$ is the $H$-band luminosity estimated as $\log L_{H} = 11.36 - 0.5 \times H + 2 \times \log D$ (in solar units) where $D$ is the distance to the source (in Mpc), $M_{tot}$ is the dynamical mass at the B-band 25~mag~arcsec$^{-2}$ isophotal radius, $M_{tot}/L_{H}$ is taken to be constant and equal to 4.6 (in solar units) and $DM_{cont}$ is the dark matter contribution to the $M_{tot}/L_{H}$ ratio at the optical radius, assumed to be equal to 0.5.

\clearpage

\newpage

\begin{table}
\footnotesize
\begin{center}
\caption{Typical uncertainties of the NUV and FUV magnitudes as a function of the magnitude.\label{error}}
\begin{tabular}{ccc}
\tableline\tableline
AB Magnitude & $\sigma$(NUV) & $\sigma$(FUV) \\
interval & (mag)         & (mag) \\
\tableline
$\leq 16$ & 0.01 & 0.02 \\
$16 ~ ... ~ 18$ & 0.03 & 0.05 \\
$18 ~ ... ~ 20$ & 0.09 & 0.15 \\
$20 ~ ... ~ 22$ & 0.26 & 0.40 \\
\tableline
\end{tabular}
\end{center}
\end{table}
   
         
         
\begin{table}
\footnotesize
\begin{center}
\caption{Basic properties of the sample galaxies: (1) Name; (2) Flag indicating membership to the NUV-selected subsample; (3) Flag indicating membership to the FIR-selected subsample; (4) R.A. (J2000); (5) Dec. (J2000); (6) Heliocentric velocity; (7) Distance derived from the velocity corrected for the Local Group infall onto Virgo and $H_{0} = 70~km~s^{-1}~Mpc^{-1}$; (8) Morphological type from NED; (9) IRAS identificator: F for FSC origin; Q,O,R for PSCz origin; SCANPI for absence in both catalogs.\label{pro}}
\begin{tabular}{lcccccrcl}
\tableline\tableline
Name & UVsel & FIRsel & R.A. & Dec. & vel. & Dist & Type & IRAS Id.\\
 & & & \multicolumn{2}{c}{(J2000.0)} & (km~sec$^{-1}$) & (Mpc) & & \\
\tableline
             MRK~544 & Y & N &  0  4  48.70 &  $-$ 1 29  54.60 &  7110 &  101.39 &           S? &  F00022-0146 \\
              NGC~10 & Y & Y &  0  8  34.56 &  $-$33 51  27.25 &  6811 &   94.62 &          Sbc &  Q00060-3408 \\
              NGC~35 & Y & Y &  0 11  10.46 &  $-$12  1  14.74 &  5964 &   83.95 &           Sb &  Q00086-1217 \\
              NGC~47 & Y & Y &  0 14  30.42 &  $-$ 7 10   6.28 &  5700 &   80.54 &          Sbc &  Q00119-0726 \\
             NGC~101 & Y & N &  0 23  54.72 &  $-$32 32   9.06 &  3383 &   45.92 &           Sc &  F00214-3248 \\
\tableline
\end{tabular}
\end{center}
Note: The distances to UGCA~438 and UGC~12613 were taken from Karachentsev et al. (2002) and Hoessel et al. (1990) respectively.
\end{table}
   

\begin{table}
\footnotesize
\begin{center}
\caption{Photometric properties of the sample galaxies: (1) Optical identifier; (2) NUV magnitude corrected for Galactic extinction; (3) FUV magnitude corrected for Galactic extinction; (4) Flux density at 60$\mu$m in Jy; (5) Flux density at 100$\mu$m in Jy; (6) $H$ magnitude from 2MASS Extended Source Caztalog. For galaxies with no detection at 2MASS we adopt the limiting value of $H = 13.9$~mag (3~mJy) as given by Jarrett et al. (2000); (7) NUV attenuation in mag, derived as indicated in Buat et al. (2005); (8) FUV attenuation derived as indicated in Buat et al. (2005). 
For some galaxies Eqs.~\ref{attenuv_eq} and \ref{attefuv_eq} gave negative values of the NUV and FUV attenuations which is senseless. In fact, this is an artifact of the polynomial fitting used to derive a functional form for the attenuation in Buat et al. (2005). Throughout this paper they will be considered as zero.\label{photo}}
\begin{tabular}{lccccccc}
\tableline\tableline
Name & $AB_{NUV}$ & $AB_{FUV}$ & $f_{60}$ & $f_{100}$ & $H$ & $A_{NUV}$ & $A_{FUV}$ \\
 & \multicolumn{2}{c}{(mag)} & \multicolumn{2}{c}{(Jy)} & (mag) & \multicolumn{2}{c}{(mag)} \\
\tableline
             MRK~544 &     15.75 &     15.97 &       0.49 &       1.11 &      12.50 &  0.71 &  0.96 \\
              NGC~10 &     15.48 &     15.98 &       0.66 &       2.97 &       9.50 &  1.59 &  2.10 \\
              NGC~35 &     15.55 &     15.89 &       1.31 &       2.34 &      11.91 &  1.29 &  1.67 \\
              NGC~47 &     15.46 &     15.97 &       0.85 &       2.57 &      10.25 &  1.02 &  1.49 \\
             NGC~101 &     14.78 &     14.98 &       0.55 &       1.75 &      10.74 &  0.50 &  0.72 \\
\tableline
\end{tabular}
\end{center}
\end{table}


\begin{table}
\footnotesize
\begin{center}
\caption{Star formation related properties of the sample galaxies: (1) Identifier of the galaxy; (2) $SFR_{NUV}$ from Eq.~\ref{sfrnuv_eq}; (3) $SFR_{FUV}$ from Eq.~\ref{sfrfuv_eq}; (4) $SFR_{dust}$ from Eq.~\ref{sfrfir_eq}; (5) $SFR_{tot}(NUV)$ from Eq.~\ref{sfrtot_eq}; (6) $SFR_{tot}(FUV)$ from Eq.~\ref{sfrtot_eq} but using $SFR^{0}_{FUV}$ instead of $SFR^{0}_{NUV}$; (7) $\left< SFR \right>$ averaged over the galaxy's lifetime, estimated as indicated in Appendix~\ref{apendice_b}.\label{sfr_tab}}
\begin{tabular}{lcccccc}
\tableline\tableline
Name & $\log SFR_{NUV}$ & $\log SFR_{FUV}$ & $\log SFR_{dust}$ & $\log SFR_{tot}(NUV)$ & $\log SFR_{tot}(FUV)$ & $\log \left< SFR \right>$ \\
 & \multicolumn{6}{c}{($M_{\odot}$~yr$^{-1}$)} \\
\tableline
             MRK~544 &       0.83 &       0.85 &       0.60 &       0.80 &       0.75 &       0.81 \\
              NGC~10 &       1.23 &       1.24 &       1.07 &       1.09 &       1.02 &       1.95 \\
              NGC~35 &       0.98 &       1.00 &       0.78 &       0.84 &       0.79 &       0.88 \\
              NGC~47 &       0.87 &       0.86 &       0.78 &       0.85 &       0.77 &       1.51 \\
             NGC~101 &       0.45 &       0.46 &       0.13 &       0.43 &       0.38 &       0.83 \\
\tableline
\end{tabular}
\end{center}
\end{table}




\clearpage

\clearpage

\clearpage

\clearpage

\clearpage

\clearpage

\clearpage

\clearpage

\clearpage

\clearpage

\clearpage

\clearpage

\clearpage

\begin{thebibliography}{}
\bibitem[Balland et al.(2003)]{2003MNRAS.343..107B} Balland, C., Devriendt, 
J.~E.~G., \& Silk, J.\ 2003, \mnras, 343, 107 
\bibitem[Bell(2003)]{2003ApJ...586..794B} Bell, E.~F.\ 2003, \apj, 586, 794 
\bibitem[Bertin \& Arnouts(1996)]{1996A&AS..117..393B} Bertin, E.~\& Arnouts, S.\ 1996, \aaps, 117, 393 
\bibitem[Boselli, Gavazzi, Donas, \& Scodeggio(2001)]{2001AJ....121..753B} 
  Boselli, A., Gavazzi, G., Donas, J., \& Scodeggio, M.\ 2001, \aj, 121, 753 
\bibitem[Brinchmann \& Ellis(2000)]{2000ApJ...536L..77B} Brinchmann, J., \& 
Ellis, R.~S.\ 2000, \apjl, 536, L77 
\bibitem[Buat \& Xu(1996)]{1996A&A...306...61B} Buat, V., \& Xu, C.\ 1996, 
\aap, 306, 61 
\bibitem[Buat \& Burgarella(1998)]{1998A&A...334..772B} Buat, V., \& 
Burgarella, D.\ 1998, \aap, 334, 772 
\bibitem[Buat et al.(1999)]{1999A&A...352..371B} Buat, V., Donas, J., 
Milliard, B., \& Xu, C.\ 1999, \aap, 352, 371 
\bibitem[Buat et al.(2005)]{2005ApJ...619L..51B} Buat, V., et al.\ 2005, 
\apjl, 619, L51 
\bibitem[Burgarella et al.(2005)]{2005MNRAS.360.1413B} Burgarella, D., 
Buat, V., \& Iglesias-P{\'a}ramo, J.\ 2005, \mnras, 360, 1413 
\bibitem[Cardelli et al.(1989)]{1989ApJ...345..245C} Cardelli, J.~A., 
Clayton, G.~C., \& Mathis, J.~S.\ 1989, \apj, 345, 245 
\bibitem[Cardiel et al.(2003)]{2003ApJ...584...76C} Cardiel, N., Elbaz, D., 
Schiavon, R.~P., Willmer, C.~N.~A., Koo, D.~C., Phillips, A.~C., \& 
Gallego, J.\ 2003, \apj, 584, 76 
\bibitem[Charmandaris et al.(2004)]{2004ApJ...600L..15C} Charmandaris, V., 
Le Floc'h, E., \& Mirabel, I.~F.\ 2004, \apjl, 600, L15 
\bibitem[Chary \& Elbaz(2001)]{2001ApJ...556..562C} Chary, R., \& Elbaz, 
D.\ 2001, \apj, 556, 562 
\bibitem[Cowie et al.(1996)]{1996AJ....112..839C} Cowie, L.~L., Songaila, 
A., Hu, E.~M., \& Cohen, J.~G.\ 1996, \aj, 112, 839 
\bibitem[Dale et al.(2001)]{2001ApJ...549..215D} Dale, D.~A., Helou, G., 
  Contursi, A., Silbermann, N.~A., \& Kolhatkar, S.\ 2001, \apj, 549, 215 
\bibitem[Devriendt \& Guiderdoni(2000)]{2000A&A...363..851D} Devriendt, 
J.~E.~G., \& Guiderdoni, B.\ 2000, \aap, 363, 851 
\bibitem[Flores et al.(1999)]{1999ApJ...517..148F} Flores, H., et al.\ 
1999, \apj, 517, 148 
\bibitem[Garnett \& Shields(1987)]{1987ApJ...317...82G} Garnett, D.~R., \& 
Shields, G.~A.\ 1987, \apj, 317, 82 
\bibitem[Gavazzi et al.(1996)]{1996A&A...312..397G} Gavazzi, G., Pierini, 
D., \& Boselli, A.\ 1996, \aap, 312, 397 
\bibitem[Gordon et al.(1997)]{1997ApJ...487..625G} Gordon, K.~D., Calzetti, 
D., \& Witt, A.~N.\ 1997, \apj, 487, 625 
\bibitem[Gordon et al.(2000)]{2000ApJ...533..236G} Gordon, K.~D., Clayton, 
G.~C., Witt, A.~N., \& Misselt, K.~A.\ 2000, \apj, 533, 236 
\bibitem[Guiderdoni \& Rocca-Volmerange(1987)]{1987A&A...186....1G} 
Guiderdoni, B., \& Rocca-Volmerange, B.\ 1987, \aap, 186, 1 
\bibitem[Heavens et al.(2004)]{2004Natur.428..625H} Heavens, A., Panter, 
B., Jimenez, R., \& Dunlop, J.\ 2004, \nat, 428, 625 
\bibitem[Heckman et al.(2005)]{2005ApJ...619L..35H} Heckman, T.~M., et al.\ 
2005, \apjl, 619, L35 
\bibitem[Hirashita, Buat, \& Inoue (2003)]{2003A&A...410...83H} Hirashita, 
H., Buat, V., \& Inoue, A.~K.\ 2003, \aap, 410, 83 
\bibitem[Hoessel et al.(1990)]{1990AJ....100.1151H} Hoessel, J.~G., Abbott, 
M.~J., Saha, A., Mossman, A.~E., \& Danielson, G.~E.\ 1990, \aj, 100, 1151 
\bibitem[Iglesias-P{\' a}ramo et al.(2004)]{2004A&A...419..109I} 
  Iglesias-P{\' a}ramo, J., Buat, V., Donas, J., Boselli, A., \& Milliard, 
  B.\ 2004, \aap, 419, 109 
\bibitem[Jarrett et al.(2000)]{2000AJ....119.2498J} Jarrett, T.~H., 
Chester, T., Cutri, R., Schneider, S., Skrutskie, M., \& Huchra, J.~P.\ 
2000, \aj, 119, 2498 
\bibitem[Karachentsev et al.(2002)]{2002A&A...389..812K} Karachentsev, 
I.~D., et al.\ 2002, \aap, 389, 812 
\bibitem[Kennicutt(1998)]{1998ARA&A..36..189K} Kennicutt, R.~C.\ 1998, 
\araa, 36, 189 
\bibitem[Kong, Charlot, Brinchmann, \& Fall(2004)]{2004MNRAS.349..769K} 
  Kong, X., Charlot, S., Brinchmann, J., \& Fall, S.~M.\ 2004, \mnras, 349, 
  769 
\bibitem[Leitherer et al.(1999)]{1999ApJS..123....3L} Leitherer, C., et 
  al.\ 1999, \apjs, 123, 3 
\bibitem[Lilly et al.(1996)]{1996ApJ...460L...1L} Lilly, S.~J., Le Fevre, 
O., Hammer, F., \& Crampton, D.\ 1996, \apjl, 460, L1 
\bibitem[Lonsdale Persson \& Helou(1987)]{1987ApJ...314..513L} Lonsdale 
Persson, C.~J., \& Helou, G.\ 1987, \apj, 314, 513 
\bibitem[\protect\citeauthoryear{Lynden-Bell}{1971}]{lynden_bell71}
  Lynden-Bell, D.\ 1971, MNRAS, 155, 95
\bibitem[Madau et al.(1996)]{1996MNRAS.283.1388M} Madau, P., Ferguson, 
H.~C., Dickinson, M.~E., Giavalisco, M., Steidel, C.~C., \& Fruchter, A.\ 
1996, \mnras, 283, 1388 
\apjs, 107, 215 
\bibitem[Martin et al.(2005)]{2005ApJ...619L...1M} Martin, D.~C., et al.\ 
2005a, \apjl, 619, L1 
\bibitem[Martin et al.(2005)]{2005ApJ...619L..59M} Martin, D.~C., et al.\ 
2005b, \apjl, 619, L59 
\bibitem[Meurer, Heckman, \& Calzetti(1999)]{1999ApJ...521...64M}
  Meurer, G.~R., Heckman, T.~M., \& Calzetti, D.\ 1999, \apj, 521, 64 
\bibitem[Morrissey et al.(2005)]{2005ApJ...619L...7M} Morrissey, P., et 
al.\ 2005, \apjl, 619, L7 
\bibitem[Moshir et al.(1990)]{moshir90}
  Moshir, M. et al.\ 1990, IRAS Faint Source Catalogue, version 2.0
\bibitem[Panuzzo et al.(2003)]{2003A&A...409...99P} Panuzzo, P., Bressan, 
A., Granato, G.~L., Silva, L., \& Danese, L.\ 2003, \aap, 409, 99 
\bibitem[Peebles(1993)]{peebles93}
 Peebles, P.~J.~E.\ 1993, Principles of Physical Cosmology (Princeton: 
 Princeton University Press), pp.119--121
\bibitem[Rowan-Robinson et al.(1997)]{1997MNRAS.289..490R} Rowan-Robinson, 
M., et al.\ 1997, \mnras, 289, 490 
\bibitem[Salim et al.(2005)]{2005ApJ...619L..39S} Salim, S., et al.\ 2005, 
\apjl, 619, L39 
\bibitem[Saunders et al.(2000)]{saunders00}
  Saunders, W., et al.\ 2000, \mnras, 317, 55
\bibitem[Sauvage \& Thuan(1992)]{sauvage92}
  Sauvage, M., \& Thuan, T.~X.\ 1992, \apj, 396, L69
\bibitem[Scalo(1986)]{1986FCPh...11....1S} Scalo, J.~M.\ 1986, Fundamentals 
of Cosmic Physics, 11, 1 
\bibitem[Schiminovich et al.(2005)]{2005ApJ...619L..47S} Schiminovich, D., 
et al.\ 2005, \apjl, 619, L47 
\bibitem[Schlegel et al.(1998)]{1998ApJ...500..525S} Schlegel, D.~J., 
Finkbeiner, D.~P., \& Davis, M.\ 1998, \apj, 500, 525 
\bibitem[Steidel et al.(1999)]{1999ApJ...519....1S} Steidel, C.~C., 
Adelberger, K.~L., Giavalisco, M., Dickinson, M., \& Pettini, M.\ 1999, 
\apj, 519, 1 
\bibitem[Sullivan et al.(2001)]{2001ApJ...558...72S} Sullivan, M., 
Mobasher, B., Chan, B., Cram, L., Ellis, R., Treyer, M., \& Hopkins, A.\ 
2001, \apj, 558, 72 
\bibitem[\protect\citeauthoryear{Takeuchi, Yoshikawa, \& Ishii}
{Takeuchi et al.}{2000}]{takeuchi00}
  Takeuchi, T.\ T.,  Yoshikawa, K., Ishii, T.\ T.\ 2000, \apjs, 129, 1
\bibitem[\protect\citeauthoryear{Takeuchi, Yoshikawa, \& Ishii}
{Takeuchi et al.}{2003}]{takeuchi03}
  Takeuchi, T.\ T.,  Yoshikawa, K., Ishii, T.\ T.\ 2003, \apj, 587, L89
\bibitem[Takeuchi et al.(2005)]{2005A&A...440L..17T} Takeuchi, T.~T., Buat, 
V., \& Burgarella, D.\ 2005, \aap, 440, L17 
\bibitem[Vijh et al.(2003)]{2003ApJ...587..533V} Vijh, U.~P., Witt, A.~N., 
\& Gordon, K.~D.\ 2003, \apj, 587, 533 
\bibitem[Wang \& Heckman(1996)]{1996ApJ...457..645W} Wang, B., \& Heckman, 
T.~M.\ 1996, \apj, 457, 645 
\bibitem[Witt \& Gordon(2000)]{2000ApJ...528..799W} Witt, A.~N., \& Gordon, 
K.~D.\ 2000, \apj, 528, 799 
\bibitem[Wyder et al.(2005)]{2005ApJ...619L..15W} Wyder, T.~K., et al.\ 
2005, \apjl, 619, L15 
\bibitem[Zaritsky(1993)]{1993PASP..105.1006Z} Zaritsky, D.\ 1993, \pasp, 
105, 1006 
\end{thebibliography}
\end{document}